\documentclass[conference]{IEEEtran}
\IEEEoverridecommandlockouts

\usepackage{amsmath,amsthm}
\usepackage[linesnumbered,ruled]{algorithm2e}
\usepackage{algorithmic}
\usepackage{cleveref}
\usepackage{booktabs}
\usepackage{subfiles} 
\usepackage{multirow}
\usepackage{xspace}
\usepackage{cite}
\usepackage{amsmath,amssymb,amsfonts}
\usepackage{algorithmic}
\usepackage{graphicx}
\usepackage{textcomp}
\usepackage{tabularx}
\usepackage{xcolor}
\usepackage{tikz}
\usetikzlibrary{calc}
\usetikzlibrary{arrows.meta}
\usetikzlibrary{shapes}
\usetikzlibrary{arrows}
\usepackage{pgfplots}
\usepackage{pgfplotstable}
\pgfplotsset{compat=1.14}
\usepackage{caption}
\usepackage{subcaption}
\usetikzlibrary{patterns}
\usetikzlibrary{shapes.geometric}
\usepackage[shortlabels]{enumitem}
\setenumerate{1.,noitemsep,topsep=2pt,leftmargin=15pt}
\setitemize{topsep=2pt,leftmargin=15pt}
\usepackage{multicol}
\usepackage[htt]{hyphenat}
\usepackage{makecell}
\usepackage[scaled=0.88]{helvet}
    
\newtheoremstyle{abcd}
  {}
  {}
  {\itshape}
  {}
  {\bfseries}
  {.}
  {.5em}
  {}
\theoremstyle{abcd}

\newtheorem{problem}{Problem}
\SetKwFor{For}{for}{do}{end\ for for}%
\definecolor{Ours0Color}{HTML}{ABDDA4}
\definecolor{Ours16Color}{HTML}{72C166}
\definecolor{Ours32Color}{HTML}{38A528}
\definecolor{PrestoColor}{HTML}{999999}
\definecolor{PostgresColor}{HTML}{6D6D6D}
\definecolor{GreenColor}{HTML}{38A528}
\definecolor{YellowColor}{HTML}{ffb570}
\definecolor{BlueColor}{HTML}{7081ff}
\definecolor{PinkColor}{HTML}{ffb0c2}

\definecolor{ComputeColor}{HTML}{ffb0c2}
\definecolor{ReadColor}{HTML}{cf3457}
\definecolor{WriteColor}{HTML}{ffb570}

\definecolor{BaseColor}{HTML}{ABDDA4}
\definecolor{OursColor}{HTML}{38A528}
\definecolor{GreedyColor}{HTML}{7081ff}
\definecolor{RandomColor}{HTML}{ffb570}
\definecolor{NoneColor}{HTML}{6D6D6D}

\definecolor{Redborder}{HTML}{805861}
\definecolor{Greenborder}{HTML}{384180}
\definecolor{Blueborder}{HTML}{566F52}
\definecolor{Greyborder}{HTML}{4D4D4D}

\definecolor{FlagColor}{HTML}{CCCCCC}

\definecolor{ExampleColor1}{HTML}{7081ff}
\definecolor{ExampleColor2}{HTML}{ffb0c2}

\definecolor{NoOptColor}{HTML}{264653}
\definecolor{LRUColor}{HTML}{777777}
\definecolor{RandomColor}{HTML}{2a9d8f}
\definecolor{GreedyColor}{HTML}{e9c46a}
\definecolor{HeuristicColor}{HTML}{f4a261}
\definecolor{SCColor}{HTML}{e76f51}
\definecolor{AllColor}{HTML}{CCCCCC}

\definecolor{SAColor}{HTML}{ffb0c2}
\definecolor{SeparatorColor}{HTML}{9b5de5}
\definecolor{BlueColor}{HTML}{0081a7}

\newcommand{\ignore}[1]{}

\newcommand{\system}{{\sf S/C}\xspace}
\newcommand{\memory}{Memory Catalog\xspace}

\newcommand{\optimizationproblem}{{\sf S/C Opt}\xspace}
\newcommand{\subproblemnodes}{{\sf S/C Opt Nodes}\xspace}
\newcommand{\subproblemorder}{{\sf S/C Opt Order}\xspace}
\newcommand{\solutionnodes}{{\sf MKP}\xspace}
\newcommand{\solutionorder}{{\sf MA-DFS}\xspace}
\newcommand{\greedy}{{\textsf Greedy}\xspace}
\newcommand{\random}{{\sf Random}\xspace}
\newcommand{\simulatedannealing}{{\sf SA}\xspace}

\newcommand{\ratio}{{\sf Ratio-based selection}\xspace}
\newcommand{\ratioshort}{{\sf Ratio}\xspace}
\newcommand{\separator}{{\sf Separator}\xspace}

\newcommand{\graph}{{\sf ExecutionGraph}\xspace}
\newcommand{\node}{{\sf ExecutionNode}\xspace}
\newcommand{\tpcds}{{\sf TPC-DS}\xspace}
\newcommand{\tpcdsdate}{{\sf TPC-DSp}\xspace}

\crefname{appendix}{Appendix}{Appendices}
\Crefname{appendix}{Appendix}{Appendices}
\crefname{figure}{Figure}{Figures}
\Crefname{figure}{Figure}{Figures}
\crefname{equation}{Equation}{Equations}
\Crefname{equation}{Equation}{Equations}
\crefname{table}{Table}{Tables}
\Crefname{table}{Table}{Tables}
\crefname{lemma}{Lemma}{Lemmas}
\Crefname{lemma}{Lemma}{Lemmas}
\crefname{theorem}{Theorem}{Theorems}
\Crefname{theorem}{Theorem}{Theorems}
\crefname{problem}{Problem}{Problems}
\Crefname{problem}{Problem}{Problems}
\crefname{algorithm}{Algorithm}{Algorithms}
\Crefname{algorithm}{Algorithm}{Algorithms}
\crefname{section}{§}{§§}
\Crefname{section}{§}{§§}

\def\BibTeX{{\rm B\kern-.05em{\sc i\kern-.025em b}\kern-.08em
    T\kern-.1667em\lower.7ex\hbox{E}\kern-.125emX}}

\makeatletter
\def\@IEEEsectpunct{.\ \,}
\def\paragraph{\@startsection{paragraph}{4}{\z@}{1.5ex plus 1.5ex minus 0.5ex}%
{0ex}{\it\normalsize\sffamily\bfseries}}
\makeatother

\newcommand{\mypara}[1]{\paragraph*{#1}}
\begin{document}
\setlength{\abovedisplayskip}{4pt}
\setlength{\belowdisplayskip}{4pt}
\title{\system: Speeding up Data Materialization with Bounded Memory}

\author{\IEEEauthorblockN{Zhaoheng Li, Xinyu Pi, Yongjoo Park}
\IEEEauthorblockA{
\textit{CreateLab @UIUC}\\
\{zl20, xinyupi2, yongjoo\}@illinois.edu}
}

\maketitle
\begin{abstract}

With data pipeline tools and the expressiveness of SQL,
    managing 
        interdependent materialized views (MVs)
    are becoming increasingly easy.
These MVs 
        are updated repeatedly upon new data ingestion (e.g., daily),
        from which database admins can observe performance metrics
    (e.g., refresh time of each MV, size on disk)
    in a consistent way for different types of updates (full vs.~incremental)
        and for different systems (single node, distributed, cloud-hosted).
One missed opportunity is that
    existing data systems treat those MV updates 
        as independent SQL statements
            without fully exploiting their dependency information and performance metrics.
However,
    if we know that the result of a SQL statement will
        be consumed immediately after for subsequent operations,
    those subsequent operations do not have to wait until the early results
        are fully materialized on storage
    because the results are already readily available in memory.
Of course, this may come at a cost
    because keeping those results in memory (even temporarily)
        will reduce the amount of available memory;
    thus, our decision should be careful.

In this paper, we introduce a new system, called \system, which tackles this problem
  through \emph{efficient creation and update of a set of MVs with acyclic dependencies among them.}
\system judiciously uses bounded memory
    to reduce the end-to-end MV refresh time by
short-circuiting expensive reads and writes;
\system's objective function accurately estimates 
the time savings from
keeping intermediate data in memory for particular periods.
Our solution jointly optimizes an MV refresh order, 
what data to keep in memory, 
and when to release the data from memory.
At a high level, \system still materializes all data exactly as defined in MV definitions; 
thus, it does not impact any service-level agreements.
In our experiments with TPC-DS datasets (up to 1TB), 
we show that \system's optimization can speedup end-to-end runtime by 1.04$\times$--5.08$\times$
with (only) 1.6GB memory.

\end{abstract}



\section{Introduction}

To accelerate query processing,
        intermediate tables and (materialized) views---both of which we refer to as MVs---are 
        frequently used
        in databases~\cite{ahmad2012dbtoaster,mcsherry2013differential, zeng2016iolap, braun2015analytics, golab2009scheduling}.
When there are dependencies among MVs,
    updates follow their topological order
        to minimize redundant computation,
    which can be managed using tools
        like Google Napa~\cite{agiwal2021napa}, 
        dbt~\cite{dbt}, Apache Airflow~\cite{airflow}, LookML~\cite{lookml}, etc.
Exploiting the widespread adoption and expressive semantics of SQL,
    we can define both full refreshes
        and incremental updates~\cite{ahmad2012dbtoaster, krueger2011fast}
    for various data systems,
        ranging from
        a single-node database (e.g., PostgreSQL~\cite{postgresql}), 
        to self-managed distributed data warehouses (e.g., Hive~\cite{hive}, SparkSQL~\cite{spark-sql}, Presto~\cite{prestosql}),
        and to cloud-hosted data services (e.g., Snowflake~\cite{dageville2016snowflake}, Azure SQL~\cite{azure}).
Database admins can obtain empirical performance metrics (e.g., elapsed time for updating each MV, the size of each MV) based on the recurrent runs of update pipelines,
while the dependency relationships 
    can easily be extracted from MV definitions.

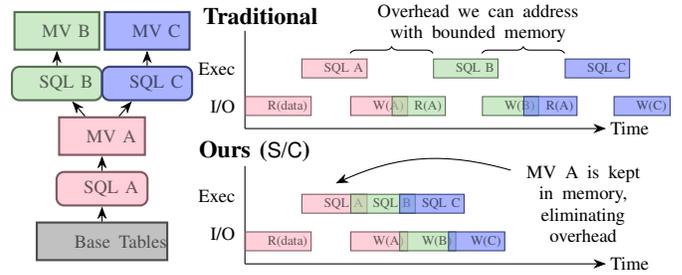
\begin{figure}[t]
\begin{subfigure}[b]{\linewidth}
\centering
\begin{tikzpicture}
\begin{scope}[every node/.style={rectangle,thick, minimum width = 1cm, minimum height = 0.5cm,fill opacity = 0.6,text opacity=1}]
    \node[text width=1.5cm, align=center, fill=PrestoColor,draw=Greyborder] (X) at (-0.4,0) {\scriptsize Base Tables};
    \node[fill=ExampleColor2,draw=Redborder] (A1) at (-0.4, 1.4) {\scriptsize MV A};
    \node[fill=BaseColor,draw=Blueborder] (B1) at (-1,2.8) {\scriptsize MV B};
    \node[fill=ExampleColor1,draw=Greenborder] (C1) at (0.2,2.8) {\scriptsize MV C} ;
\end{scope}

\begin{scope}[every node/.style={rectangle,thick, rounded corners=0.1cm,fill opacity = 0.6,text opacity=1}]
    \node[fill=ExampleColor2,draw=Redborder] (A2) at (-0.4, 0.7) {\scriptsize SQL A};
    \node[fill=BaseColor,draw=Blueborder] (B2) at (-1,2.1) {\scriptsize SQL B};
    \node[fill=ExampleColor1,draw=Greenborder] (C2) at (0.2,2.1) {\scriptsize SQL C} ;
\end{scope}

\begin{scope}[>={Stealth[black]},
              every node/.style={fill=none,circle},
              every edge/.style={draw=black}]
    \path [->] (X) edge node {} (A2);
    \path [->] (A2) edge node {} (A1);
    \path [->] (A1) edge node {} (B2);
    \path [->] (A1) edge node {} (C2);
    \path [->] (B2) edge node {} (B1);
    \path [->] (C2) edge node {} (C1);
\end{scope}

\begin{scope}[every node/.style={rectangle,thick, anchor=east}]
    \node[align=right] (label1) at (2.6, 3.0) {\small \textbf{Traditional}};
    \node[align=right] (label2) at (1.5, 2.3) {\scriptsize Exec};
    \node[align=right] (label3) at (1.5, 1.8) {\scriptsize I/O} ;
    \node[align=right] (label4) at (2.5, 1.2) {\small \textbf{Ours (\system)}};
    \node[align=right] (label5) at (1.5, 0.6) {\scriptsize Exec};
    \node[align=right] (label6) at (1.5, 0.1) {\scriptsize I/O} ;
    \node[align=right] (label7) at (7, 1.5) {\scriptsize Time};
    \node[align=right] (label8) at (7, -0.3) {\scriptsize Time} ;
\end{scope}

\begin{scope}[>={Stealth[black]},
              every node/.style={fill=none,circle},
              every edge/.style={draw=black}]
    \draw (1.5,2.8) -- (1.5, 1.5);
    \draw (1.5,1.0) -- (1.5, -0.3);
    \draw[->] (1.5,1.5) -- (6.3, 1.5);
    \draw[->] (1.5,-0.3) -- (6.3, -0.3);
\end{scope}

\begin{scope}[every node/.style={rectangle, anchor=west, inner sep=0.05cm,fill opacity = 0.6,text opacity=1}]
    \node[fill=ExampleColor2,draw=Redborder] (X) at (1.5,1.8) {\tiny R(data)};
    \node[fill=ExampleColor2,draw=Redborder] (A1) at (2.25, 2.3) {\tiny SQL A};
    \node[fill=ExampleColor2,draw=Redborder] (B1) at (2.9, 1.8) {\tiny W(A)};
    \node[fill=BaseColor,draw=Blueborder] (C1) at (3.45, 1.8) {\tiny R(A)};
    \node[fill=BaseColor,draw=Blueborder] (A1) at (4, 2.3) {\tiny SQL B};
    \node[fill=BaseColor,draw=Blueborder] (B1) at (4.65, 1.8) {\tiny W(B)};
    \node[fill=ExampleColor1,draw=Greenborder] (C1) at (5.2, 1.8) {\tiny R(A)};
    \node[fill=ExampleColor1,draw=Greenborder] (A1) at (5.75, 2.3) {\tiny SQL C};
    \node[fill=ExampleColor1,draw=Greenborder] (C1) at (6.4, 1.8) {\tiny W(C)};
\end{scope}

\begin{scope}[>={Stealth[black]},
              every node/.style={fill=none,circle},
              every edge/.style={draw=black}]
    \draw [decorate,
    decoration = {brace}] (2.9, 2.5) --  (4, 2.5);
    \draw [decorate,
    decoration = {brace}] (4.65, 2.5) --  (5.75, 2.5);
\end{scope}
\node[text width=3cm, align=center] (X) at (4.6,2.9) {\scriptsize Overhead we can address with bounded memory \par};

\begin{scope}[every node/.style={rectangle, anchor=west, inner sep=0.05cm,fill opacity = 0.6,text opacity=1}]
    \node[fill=ExampleColor2,draw=Redborder] (X) at (1.5,0.0) {\tiny R(data)};
    \node[fill=ExampleColor2,draw=Redborder] (A1) at (2.25, 0.5) {\tiny SQL A};
    \node[fill=ExampleColor2,draw=Redborder] (B1) at (2.9, 0.0) {\tiny W(A)};
    \node[fill=BaseColor,draw=Blueborder] (A1) at (2.9, 0.5) {\tiny SQL B};
    \node[fill=BaseColor,draw=Blueborder] (B1) at (3.55, 0.0) {\tiny W(B)};
    \node[fill=ExampleColor1,draw=Greenborder] (A1) at (3.55, 0.5) {\tiny SQL C};
    \node[fill=ExampleColor1,draw=Greenborder] (C1) at (4.2, 0.0) {\tiny W(C)};
\end{scope}

\begin{scope}[>={Stealth[black]},
              every node/.style={fill=none,circle},
              every edge/.style={draw=black}]
    \node[] (y) at (5.2, 0.9) {};
    \node[] (z) at (2.55, 0.7) {};
    \draw[->] (y)  to [bend right = 20] (z);
\end{scope}
\node[text width=2.1cm, align=center] (X) at (6.0,0.5) 
{\scriptsize MV A is kept in memory, eliminating overhead\par};

\end{tikzpicture}
\end{subfigure}
\vspace{-6mm}
\caption{Our intuition:
we can reduce an overall MV refresh time by exploiting existing dependency relationships.
If we know SQL B will depend on the output of SQL A,
    we don't have to wait until the result of SQL A is fully materialized.}
\vspace{-4mm}
\label{fig:intro}
\end{figure}
\begin{table*}[t]
\centering
\caption{Comparison between \system and other data warehouse optimization problems}
\label{tbl:existing_work}

\vspace{-2mm}
\small
\begin{tabular}{lll}
\toprule
\textbf{Topics}              &  \textbf{Optimization Objective}           \\ \midrule
Incremental Query Processing~\cite{tang2019intermittent, wangtempura}  & Early querying on incomplete data, progressive update with the arrival of remaining data \\

Cache prefetching~\cite{jalaparti2018netco, yang2017mithril} & SLO-aware prefetching for anticipated jobs   \\

Incremental view maintenance~\cite{griffin1995incremental,palpanas2002incremental}   
   
  & Individual MVs update faster  \\ 
 
MV refresh scheduling~\cite{golab2009scheduling,folkert2005optimizing, ahmed2020automated} 
  
  & Updates some MVs earlier than others (but the total time remains the same)    \\    
Intermediate result reuse~\cite{yang2018intermediate, dursun2017revisiting, michiardi2019memory} & Cache and reuse intermediate results to speedup future jobs (but no job reordering)\\
\textbf{Ours (\system)} 

& \textbf{Speedup end-to-end refresh time of a set of MVs via job reordering and caching data} \\
\bottomrule
\end{tabular}
\vspace{-4mm}
\end{table*}

\mypara{Missed Opportunity}

When updating a set of MVs
    that are dependent on one another,
    existing data systems do not introduce special optimizations 
        besides caching.
Our idea is that
    by exploiting (i) the explicit knowledge of dependency relationships among MVs
        and (ii) the performance metrics observed from previous runs,
    we can further improve data systems operations
        to reduce the end-to-end MV refresh time.
Among possible ways,
    we pursue one significant optimization opportunity
        based on the following observation:
\emph{While persisting intermediate data might be unavoidable due to service level agreements (SLAs), writing/reading intermediate data can be short-circuited via bounded memory (if the data fits) 
by letting a data system read input data directly from in-memory objects rather 
    then reading from persisted tables.}
\Cref{fig:intro} depicts our idea: by keeping \textsf{Table A} in memory after computation, the downstream tasks \textsf{SQL B} and \textsf{SQL C} can be executed with minimal delay in parallel with the materialization of \textsf{Table A}.
Since the memory size is finite,
    the intermediate data must be kept in memory
        only when the benefits of doing so are significant.
This approach is general because 
    we make no assumptions about the content of computation tasks 
    (i.e., full vs. incremental MV refresh) and 
    how they are performed (i.e., single-node vs. distributed).
Moreover, since some data systems 
    allow explicit data placement (e.g., Presto~\cite{prestosql}, Spark~\cite{spark-sql}, Polars~\cite{polars}),
    we can easily implement this optimization
        using existing  features
    without modifying their internals.
Note that our approach still persists all intermediate tables;
    we only reduce the wait times for subsequent updates
        exploiting the dependency relationships.

\mypara{Challenge} 

Unfortunately,
    achieving our goal is technically challenging
        because na\"ive approaches (e.g., greedily caching as much data as possible and LRU for cache eviction) offer limited performance benefits (\cref{sec:experiments}).
For significant gains,
    we must consider several factors simultaneously.
First, we must optimize the order of MV refresh, as it impacts which intermediate data we can keep and release from memory.
Off-the-shelf topological sort algorithms~\cite{networkxbfs} may not produce an order optimized for 
our problem.
Second, we must determine which intermediate MVs to keep in bounded memory.
A greedy approach (i.e., keep if there is available space) or a random selection performs poorly, according to our study.
Third, we must develop an adaptable strategy based on user workloads and intermediate data sizes; a fixed, heuristic strategy may result in suboptimal solutions if users' workloads change.
To overcome these challenges,
we pursue a principled method by
    representing MVs using a graph
        and by investigating several optimization algorithms, as will be described below.

\mypara{Our Approach}

In this paper, 
    we introduce a system (called Short-Circuit or S/C) that specializes in refreshing a graph of MVs.
S/C speeds up end-to-end MV refresh using bounded memory by exploiting the dependency relationships among MVs and their observed performance metrics.
The dependency relationships among MVs are modeled using an acyclic graph where nodes represent individual MV updates, and the edges represent dependencies.
To optimize the end-to-end performance, \system solves the following problem at a high level:
\begin{equation}
\min_{O,\, U}  \; \sum_{n_i \in O} \text{data-access-time}(n_i; U),
\label{sec:intro:objective}
\end{equation}
Where $O$ is an MV refresh order, $U$ is the set of nodes for which we temporarily keep their results in memory, and $n_i$ is the $i$-th node. 
$U$ must be \emph{feasible}; that is,
    the sum of all the intermediate data we keep in memory
        cannot exceed a predefined limit
            at any point in time while updating MVs\label{sec:intro_definition}.

To solve the above problem,
    we alternatively optimize $O$ and $U$, as follows:
    first, we find $U$ that 
        can minimize the objective function (\cref{sec:intro:objective}) for fixed $O$;
    second, we find $O$ that
        can lower overall memory usage,
            expecting
                such an order can lead to more optimal $U$ in the following iteration (\cref{sec:joint_optimization}). 
While this approach may find locally optimal solutions,
    our empirical study suggests that 
        starting from a specialized designed variant of DFS (depth-first search)
            can offer high-quality solutions (\cref{sec:exp_algm}).

\mypara{Comparison to Other Work}

This work adopts mathematical optimization 
for memory management and scheduling in data systems with a focus on reducing an end-to-end MV refresh time by exploiting observed performance data.
While optimizations and statistical techniques for data systems have been explored in various related problems (e.g., incremental query processing~\cite{tang2019intermittent,wangtempura,deepola}, cache prefetching~\cite{jalaparti2018netco, yang2017mithril}, indexing~\cite{chockchowwat2022airphant, chockchowwat2022automatically}), they are aimed towards resource management for standalone queries.
The problem addressed in this work is significantly different, notably including the joint scheduling of MV updates.
While optimizing MV updates via incremental updates~\cite{griffin1995incremental,palpanas2002incremental} and scheduling~\cite{golab2009scheduling,folkert2005optimizing, ahmed2020automated}
    pursues the same high-level goal as ours 
        because we aim to improve data preparation stages,
    the specific problem we  address in this work
        is largely orthogonal to them.
Works in intermediate result reuse~\cite{yang2018intermediate, dursun2017revisiting, michiardi2019memory, park2017database} similarly selectively store items to speedup future workloads (i.e., MV refresh runs); however, to the best of our knowledge, we are the first work that considers both job execution (re)order as well as intermediate result caching with a bounded amount of memory.
\Cref{tbl:existing_work} summarizes core differences.

Our technical contributions
    can potentially be applied to accelerating 
        a wider class of workloads
that consist of repetitive jobs with their dependencies 
    expressed in a directed acyclic graph,
such as Extract-Load-Transform (ETL) with Hadoop and Spark~\cite{spark-sql},
    job coordination via Apache Airflow~\cite{airflow} and Apache Oozie~\cite{apacheoozie},
        etc.,
However, for concrete discussions, we focus on MV refresh in this work.

\mypara{Contributions}

Our contributions are as follows:
\begin{itemize}
    \item \textbf{System architecture.} We propose the architecture of a system (\system) aimed at reducing end-to-end runtimes of MV refresh workloads in data warehouses and provide an efficient Presto-based~\cite{prestosql} implementation. (\cref{sec:overview})

    \item \textbf{Problem definition.} We propose and formally define an optimization problem (\optimizationproblem) on MV refresh workloads, namely selecting intermediate data to 
        persist in bounded memory 
        to maximize read/write time reductions. (\cref{sec:problem_setup})
    
    \item \textbf{Algorithm and Analysis.} We employ an alternating optimization algorithm to solve \optimizationproblem.
    This algorithm decomposes \optimizationproblem into two subproblems, for which we provide efficient and empirically effective solutions. (\cref{sec:joint_optimization})
    
    \item \textbf{Experimental Analysis.} 
    We verify the effectiveness of \system on improving MV refresh speed. We find that \system speedups end-to-end runtimes by 1.04$\times$--5.08$\times$
    using 1.6GB \memory on TPC-DS datasets (up to 1TB).
\end{itemize}
\section{Materialization in Modern Data Warehouses}
\label{sec:background}

Provisioning materialized views (i.e., data materialization)
is effective in speeding up OLAP queries.
In this section, we describe significant occurrences of data materialization in modern data warehouses and the I/O overhead we observe.

\subsection{Acceleration by Data Materialization}

Modern data warehouses manage massive volumes of data.
Horizontal scaling might not be cost-efficient in reducing the latencies of analytical queries.
For a better cost-benefit trade-off, data materialization is frequently employed.

\mypara{Large Data, Complex Queries}

Real-world data warehouses manage a large volume of data.
According to an analysis of 1.54 billion queries and 1.7 million tables across 40 different
data warehouse accounts~\cite{aleyasen2022overcoming, aleyasen2022intelligent},
12.1\% of tables are larger than 3.625 TB, and 29.1\% of the tables are between 725 GB and 3.625 TB.
Most analytical queries include data-intensive, time-consuming operations:
31\% of \texttt{select} queries are reported to include at least one join.
32\% of time is spent on queries longer than 10 minutes.
According to another analysis by Tableau~\cite{vogelsgesang2018get}, machine-generated queries are often complex
due to the impedance mismatch between BI-driven analysis and SQL-based expressions.
Generally, a large amount of data and costly operations lead to longer query latencies.

While horizontal scaling~\cite{spark-sql,prestosql,shvachko2010hadoop} and aggressive caching~\cite{li2014tachyon}
could mitigate the issue to some extent, they are not budget-friendly solutions.
Specifically, to benefit from caching, users must pay more to keep clusters up and running even when query workloads are relatively lightweight~\cite{gupta2015amazon}.
To save cost, one could elastically scale clusters
according to their query workloads; however, this approach increases 
the chance of cache misses, requiring data transfers over the network~\cite{dageville2016snowflake}. Fetching data over the network is slower than
reading from local storage.
Tiered storage layers~\cite{wu2019autoscaling,li2019cloud,herodotou2021trident} might be a more cost-effective solution to keep data access fast;
however, quickly fetching terabytes of data remains challenging.


\begin{figure}[t]

\centering

\begin{tikzpicture}

\begin{axis}[
    ybar stacked,
    xtick=data,
    width=75mm,
    height=32mm,
    bar width=3mm,
    ymin=0,
    ymax=100,
    axis y line*=none,
    axis x line*=none,
    xtick={1,2,...,10},
    xticklabels = {\texttt{W1}, \texttt{W2}, \texttt{W3}, \texttt{W4}, \texttt{W5}, \texttt{W6}, \texttt{W7}, \texttt{W8}, \texttt{W9}, \texttt{W10}},
    ytick={0, 20, 40, 60, 80, 100},
    xmin = 0,
    xmax = 11,
    tick label style={font=\footnotesize},
    legend style={
        at={(-0.2,1.1)},anchor=south west,column sep=2pt,
        draw=black,fill=none,line width=.5pt,
        /tikz/every even column/.append style={column sep=5pt},
        font=\footnotesize,
    },
    legend cell align={left},
    legend columns=4,
    label style={font=\footnotesize},
     xlabel={Data Warehouse Workloads},
     xlabel style={yshift = 1ex},
    ylabel={Percentage (\%)},
    ymajorgrids,
    area legend,
    legend image code/.code={%
    \draw[#1, draw=none] (0cm,-0.1cm) rectangle (0.6cm,0.1cm);}
]

\addplot[fill=BlueColor,draw=none]
table[x=x,y=y] {
x y
1 11
2 16
3 29
4 19
5 7
6 38
7 2
8 4
9 8
10 10
};

\addplot[fill=YellowColor,draw=none]
table[x=x,y=y] {
x y
1 43
2 12
3 59
4 30
5 40
6 17
7 16
8 4
9 63
10 29
};

\addplot[fill=GreenColor,draw=none]
table[x=x,y=y] {
x y
1 36
2 22
3 5
4 35
5 37
6 27
7 51
8 50
9 9
10 25
};

\addplot[fill=PinkColor,draw=none]
table[x=x,y=y] {
x y
1 10
2 50
3 7
4 16
5 16
6 18
7 31
8 42
9 20
10 36
};

\addlegendentry{Transformation}
\addlegendentry{Analytics}
\addlegendentry{Insert}
\addlegendentry{Others}

\end{axis}
\end{tikzpicture}

\vspace{-2mm}
\caption{Runtime breakdown by query type for 10 real-world workloads (\texttt{W1}--\texttt{W10}).
Data from Amirhossein et al.~\cite{aleyasen2022overcoming, aleyasen2022intelligent}.}
\label{fig:overview:workload}
\vspace{-2mm}
\end{figure}
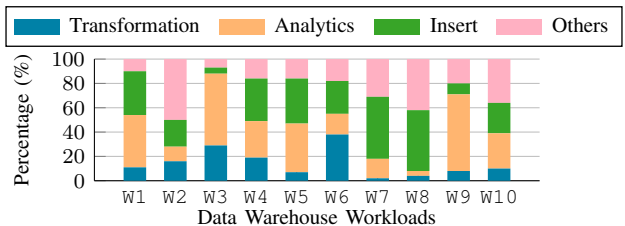

\mypara{Materialization for Faster Querying}

Data materialization can reduce query latencies by precomputing costly operations in advance
and using the results.
MVs have been studied extensively in the literature.
The topics include efficient refresh/updates of MVs~\cite{ahmad2012dbtoaster,mcsherry2013differential,armbrust2013generalized,zeng2016iolap},
recommending useful MVs for observed query workloads~\cite{shukla1998materialized,agrawal2000automated,park2017database},
optimizing queries using MVs~\cite{chaudhuri1995optimizing,calvanese2012view},
approximate aggregation~\cite{heule2013hyperloglog,flajolet2007hyperloglog,rong2020approximate}, etc.
While sketching~\cite{bigquery} and sampling~\cite{zeng2016iolap} can also be
considered as a special case of MVs, we do not discuss them.

Automated management of materialized data is gaining interest.
Napa~\cite{agiwal2021napa} enables both high-throughput data ingestion and fast query speed using MVs.
Keebo~\cite{keebo} automates materialized view creation by mining from query workloads.
Also, there are tools that can help users manually define data materialization strategies; 
we describe a few commonly used tools in the following section.

\subsection{Data Materialization: Significant Fraction}

Real-world data warehouses spend much time materializing data.
By optimizing MV refresh, we can reduce 
infrastructure costs and data staleness.
Managing complex MVs
is becoming easier due to newly available tools providing
templating languages, visualizations, and web-based interfaces.

\begin{figure}[t]

\centering

\begin{tikzpicture}

\begin{axis}[
    ybar stacked,
    xtick=data,
    width=80mm,
    height=32mm,
    bar width=4mm,
    ymin=0,
    ymax=100,
    axis y line*=none,
    axis x line*=none,
    xtick={1,2,3,4},
    xticklabels = {1G (5.4s),10G (14s),100G (52s),1000G (560s)},
    ytick={0, 20, ..., 100},
    xlabel=Data size (total processing time in seconds),
    xlabel style={yshift = 1ex},
    xmin = 0.5,
    xmax = 4.5,
    tick label style={font=\footnotesize},
    legend style={
        at={(-0.2,1.1)},anchor=south west,column sep=2pt,
        draw=black,fill=none,line width=.5pt,
        /tikz/every even column/.append style={column sep=5pt},
        font=\scriptsize,
    },
    legend cell align={left},
    legend columns=4,
    label style={font=\footnotesize},
    ylabel={Percentage (\%)},
    ymajorgrids,
    area legend,
    legend image code/.code={%
    \draw[#1, draw=none] (0cm,-0.1cm) rectangle (0.6cm,0.1cm);}
]

\addplot[fill=BlueColor,draw=none]
table[x=x,y=y] {
x y
1 5.1
2 10.4
3 11.5
4 9.7
};

\addplot[fill=YellowColor,draw=none]
table[x=x,y=y] {
x y
1 58.1
2 27.7
3 18.6
4 21.9
};

\addplot[fill=GreenColor,draw=none]
table[x=x,y=y] {
x y
1 36.8
2 61.9
3 69.9
4 68.4
};

\addlegendentry{Read (base tables)}
\addlegendentry{Compute (joins)}
\addlegendentry{Write (final output)}

\end{axis}
\end{tikzpicture}

\vspace{-2mm}
\caption{Runtime breakdown by operation. 
Compressing/writing the result to storage 
takes much longer.
A cloud data warehouse is used to join four tables 
(\texttt{customer}, \texttt{orders}, \texttt{lineitem}, \texttt{nation}) 
appearing in TPC-H Q8.}
\label{fig:overview:overhead}
\vspace{-2mm}
\end{figure}
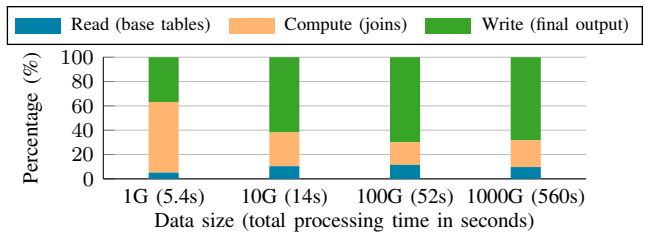

\mypara{Heavily Used in Warehouses}

Data materialization makes up a significant fraction of 
modern data warehouse workloads.
\cref{fig:overview:workload} shows statistics across ten independent
workloads~\cite{aleyasen2022overcoming}.
Based on query runtime, data materialization (e.g., DDL, DML) takes 2\%-38\% of each workload.
Interestingly, in \texttt{W6}, a warehouse spends 2.2$\times$ 
 time on materialization than analytics.

We have also spoken with several data engineers
at big and small companies about their data ingestion and preparation pipelines.
They are customers of cloud-managed data warehouses.
In their environments, periodic (e.g., daily) data ingestion (using tools like Airflow~\cite{airflow}, Fivetran~\cite{fivetran}, etc.)
is followed by several hours of recurring in-warehouse operations for
creating and updating dependent tables.\footnote{MVs are not currently first-choice because popular data warehouses like Amazon Redshift, BigQuery, and Snowflake do not support MVs with joins.}
Dashboards are set up to use these precomputed tables to serve user-facing queries (instead of computing all the way from base tables); 
thus, end-users can see results more quickly (e.g., when they change filtering conditions).
Data materialization is mostly managed via web-based tools 
(rather than hand-rewritten DDL/DML statements sent directly to warehouses), 
which we will describe shortly.

\mypara{Tools for Materialization}

With high-level admin tools, 
managing a graph of MVs has become easier.
First, Looker offers a data modeling language, LookML~\cite{lookml}. With LookML, users can express complex joins and nested structures.
Based on LookML, Looker provides a feature called PDT~\cite{looker-pdt} to 
store views as materialized tables. 
Once built, Looker uses materialized tables in its generated queries.
Second, dbt~\cite{dbt} is a data modeling tool based on nested table/view definitions. Since a definition can reference other definitions, the overall structure forms a directed acyclic graph where nodes represent table/view definitions, and edges represent the dependencies among those definitions.
dbt then uses a topological sort to determine the order of table updates.
Note that while these tools offer convenient ways of constructing
intermediate tables, all the operations are eventually compiled into SQL
and sent to connected warehouses; data warehouses then perform the
actual computation and data manipulation.
Thus, what we discuss in this paper---speeding up warehouses' internal
operations---is orthogonal to what these tools offer.

\begin{figure}[t]
\begin{subfigure}[b]{\linewidth}
\centering
\begin{tikzpicture}
 \node(mvdefinition)[minimum height=10mm, minimum width=31mm, anchor=south] at (-2.8, 0){};
  
   \node[anchor=north west, align=left] at ($(mvdefinition.north west) + (0, 0.05)$) {\small \texttt{CREATE MV \textbf{MV1}}\\[-0.2em]\small\texttt{... FROM TABLE;} \\[-0.2em]\small \texttt{CREATE MV \textbf{MV2}}\\[-0.2em]\small\texttt{...FROM \textbf{MV1};}\\[-0.2em]\small \texttt{CREATE MV \textbf{MV3}}\\[-0.2em]\small\texttt{...FROM \textbf{MV1};}};
\begin{scope}[every node/.style={rectangle,thick, minimum width = 1cm, minimum height = 0.5cm,fill opacity = 0.6,text opacity=1}]
    \node[dashed, text width=1.5cm, align=center,draw=black] (X) at (0,-0.8) {\scriptsize \texttt{TABLE}};
    \node[draw=black] (A1) at (0, 0) {\scriptsize \texttt{MV1}};
    \node[draw=black] (B1) at (-0.6,0.8) {\scriptsize \texttt{MV2}};
    \node[draw=black] (C1) at (0.6,0.8) {\scriptsize \texttt{MV3}} ;
\end{scope}

\begin{scope}[>={Stealth[black]},
              every node/.style={fill=none,circle},
              every edge/.style={draw=black}]
    \path [->] (X) edge node {} (A1);
    \path [->] (A1) edge node {} (B1);
    \path [->] (A1) edge node {} (C1);
\end{scope}


\begin{scope}[every node/.style={rectangle,thick, minimum width = 1cm, minimum height = 0.5cm,fill opacity = 0.6,text opacity=1}]
    \node[fill=PrestoColor,draw=Greyborder] (A11) at (2.9, -0.2) {\scriptsize \texttt{MV1}};
    \node[draw=Greyborder] (B11) at (2.3,0.6) {\scriptsize \texttt{MV2}};
    \node[draw=Greyborder] (C11) at (3.5,0.6) {\scriptsize \texttt{MV3}} ;
\end{scope}

\begin{scope}[>={Stealth[black]},
              every node/.style={fill=none,circle},
              every edge/.style={draw=black}]
    \path [->] (A11) edge node {} (B11);
    \path [->] (A11) edge node {} (C11);
\end{scope}

\node[] (yy) at (3.2, -0.8) {\small Keep in memory};
\node[draw,circle, fill=FlagColor,inner sep=0.05cm,minimum width = 3mm,text opacity=1] at (1.8, -0.8) {};

\node[] () at ($(A11.east) + (0.2,0)$) {\Large 1};
\node[] () at ($(C11.east) + (0.2,0)$) {\Large 3};
\node[] () at ($(B11.west) + (-0.2,0)$) {\Large 2};


\end{tikzpicture}
\end{subfigure}
\vspace{-5mm}
\caption{Input to \system for an example workload. Left: The workload's dependency graph. Right: \textit{plan} (execution order, nodes to keep in memory) for the MV refresh run by Optimizer.}
\label{fig:input}
\vspace{-4mm}
\end{figure}
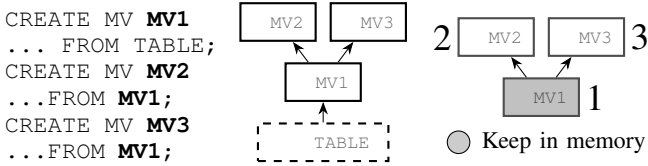
\subsection{Overhead of Data Materialization}

Reading/writing data is often a significant bottleneck in data systems.
That is, 
much of 
data transformation/materialization 
is spent on persisting data.

First, read/write often carries a large overhead (compared to compute) in modern data warehouses.
To study this, we measured how much time a system spends on 
read/write operations (e.g., serialization, compression) relative to compute operations (e.g., join, filtering).
Using an (anonymous) commercial data warehouse managed by its vendor,
we ran CTAS statements containing three inner joins (or equivalently, four joined tables) which appear in TPC-H query \#8---joins of \texttt{customer}, \texttt{orders}, \texttt{lineitem}, and \texttt{nation}.
\Cref{fig:overview:overhead} shows the result for different sizes of datasets, i.e., from scale factor 1 (or 1 GB) to scale factor 1000 (or 1000 GB). On the X-axis, a dataset size is accompanied by a respective total runtime (i.e., 5.4 seconds for the 1 GB dataset).
According to this study, writing joined results into persistent storage (which could include compression, serialization, and network I/O) took 37\%--69\% of the total runtime (of each statement).

Second, read/write also takes significant time for open-source data systems.
We compared a few different implementations for Apache Parquet (e.g., C++ Arrow~\cite{arrow-cpp}, Rust Arrow~\cite{arrow-rust}), 
a widely used columnar data format for data warehouses~\cite{spark-sql,prestosql,hive,athena,snowflake-parquet}. 
In our environment, we could achieve the best serialization speed with Rust Arrow compiled from its source.
Under this setup, we tested the overhead of read/write in creating intermediate tables for TPC-DS queries (see \cref{sec:exp:setup}).
Overall, read/write took 85\% of the time spent on compute operations (e.g., join, filtering, etc.).


\section{\system: System Overview}
\label{sec:overview}
\begin{figure}[t]
\begin{subfigure}{\columnwidth}
\centering
\begin{tikzpicture}
\tikzset{cbox/.style={
	minimum height=10mm,draw=black,fill=white,ultra thick,align=center,
	minimum width=14mm,anchor=north
}}
\tikzset{dbox/.style={
	minimum height=6mm,minimum width=14mm,draw=black,fill=white,thick,
    font=\scriptsize, align=center,anchor=north,inner sep=1mm
}}

\node(bg)
 [minimum height=32mm, minimum width=42mm,
  draw=black, fill=gray, opacity=0.2, anchor=north,
  dashed,ultra thick]
 at (2.2,0.75) {};
\node [anchor=north east,font=\bf\Large] at ($(bg.north east)$) {\system};
 
\node(opt) [cbox,minimum height = 17mm, minimum width = 17mm, align=center,anchor=center]
 at ($(bg.south east) + (-1.0, 1.05)$) {};
 
\node [anchor=center] at ($(opt.north) + (0,-1.2)$) {\bf \small Optimizer};

\node(optalgm) [dbox,
  minimum height=6mm,minimum width=10mm]
 at ($(opt.north) + (0, -0.175)$) {Optimization\\algorithm};
 
\node(metadata) [dbox,align=center,anchor=center]
 at ($(opt.north) + (0, 0.4)$) {Metadata};
 
\node(engine) [cbox,minimum height = 29mm, minimum width = 20mm, align=center,anchor=center]
 at ($(opt.west) + (-1.15, 0.6)$) {};
 
  \node(mqueue) [dbox,align=center,anchor=center]
 at ($(engine.south) + (0,0.5)$) {Materialization\\queue};
 
  \node(optresult) [dbox,align=center,anchor=south]
 at ($(mqueue.north) + (0,0.1)$) {Optimization\\result};
 
 \node(depgraph) [dbox,align=center,anchor=south]
 at ($(optresult.north) + (0,0.1)$) {Dependency\\graph};
 
 \node [anchor=south] at ($(depgraph.north) + (0,0.025)$) {\bf \small Controller};
 
\node(database) [cbox,minimum height = 22mm, minimum width = 17mm, align=center,anchor=east]
 at ($(bg.west) + (-0.2, -0.35)$) {};
 
   \node(inmemory) [dbox,align=center,anchor=center]
 at ($(database.south) + (0,0.5)$) {Memory\\ Catalog};
 
  \node(ondisk) [dbox,align=center,anchor=south]
 at ($(inmemory.north) + (0,0.1)$) {Physical\\ catalog };
 
  \node [anchor=south] at ($(ondisk.north) + (0,0.05)$) {\bf \small DBMS};

\draw[thick,->] 
($(metadata.west) + (-0.3, 0)$) --
($(metadata.west)$);

\draw[thick,->] 
($(metadata.south)$) --
($(optalgm.north)$);

\draw[thick,->] 
($(optresult.east) + (0.5, 0)$) --
($(optresult.east)$);

\draw[thick,->] 
($(optresult.west)$) --
($(optresult.west) + (-0.7, 0)$);

\draw[thick,->] 
($(mqueue.west)$) --
($(inmemory.east) + (0, 0.025)$);

\node(disk) [dbox,align=center,font=\footnotesize,minimum width=16mm,anchor=east] at ($(ondisk.west)+(-0.6,0)$) {external\\ storage};

\draw[thick,->] 
($(ondisk.west)$) --
($(disk.east)$);

\draw[thick,->] 
($(disk.east)$) --
($(ondisk.west)$);

\draw[thick,-] 
($(inmemory.west)$) --
($(disk.south) + (0, -0.425)$);

\draw[thick,->] 
($(disk.south) + (0, -0.425)$) --
($(disk.south)$);

    \node [anchor=north] at ($(disk.south) + (0.2,-0.375)$) {\footnotesize Materialization};
    
\node(input) [dbox,align=center,font=\footnotesize,anchor=south] at ($(disk.north)+(0, 0.5)$) {SQL\\ workload};

\draw[thick,->] 
($(input.east)$) --
($(input.east) + (2.5, 0)$);

\end{tikzpicture}
\label{fig:internal_architecture}
\end{subfigure}
\vspace{-5mm}
\caption{\system architecture. 
Optimizer determines which intermediate data to keep in \memory. 
DBMS can directly access data (temporarily) kept in \memory for computations, saving the time for reading data from external storage.}
\label{fig:system_overview}
\vspace{-4mm}
\end{figure}
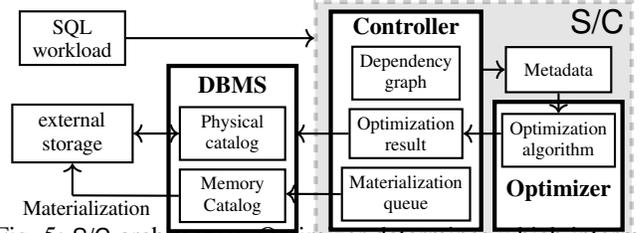

\system is a system such that, given a list of materialized view (MV) definitions, 
it refreshes all the MVs (or more accurately, it generates data for those MVs).
This section overviews how \system performs this operation: 
\cref{sec:inputs} describes \system's inputs,
\cref{sec:arch} introduces \system's subcomponents and their roles and
\cref{sec:job_representation} overviews the execution of an MV refresh run.

\subsection{Workload Specification}
\label{sec:inputs}

A list of MVs to update (and their dependencies) are formally expressed
using a \emph{dependency graph}.
Then, \emph{execution metadata}
records additional information about those MVs
    (e.g., the size on disk)
useful for optimization.

\mypara{Dependency Graph}

The set of MVs to refresh is to be provided to \system in the form of a dependency graph (\Cref{fig:input}, left). 
Each node corresponds to a single MV update, and edges correspond to dependencies between MVs.
Each node contains a SQL statement for a specific MV update.

\mypara{Execution Metadata} 

\system's optimization requires information from DBMS-side SQL executions from past MV refresh runs, namely (1) the estimated size of the output table from executing the SQL statement in each node and (2) the estimated time savings of keeping a node output in memory (see \cref{sec:problem_setup}).

\subsection{Internal Architecture}
\label{sec:arch}

\system's Controller coordinates MV updates according to
    the \emph{plan}---order of updates and in-memory caching---created 
        by Optimizer (see \Cref{fig:system_overview}).


\mypara{Controller}

Controller directs the order of node (i.e., MV) materialization and how/where to store the output of individual nodes (i.e., keep in memory vs. materialize on external storage\footnote{Our implementation of \system uses NFS for storage. This can be substituted with alternative materialization locations.}) according to the directions provided by the Optimizer (see below). The directions are compiled into a corresponding SQL statement and sent to the DBMS for execution.
The Controller additionally manages the materialization of intermediate tables kept in memory. (\Cref{sec:job_representation})

\mypara{Optimizer}

The Optimizer computes the MV refresh order (e.g., MV1, MV2, MV3) and nodes to keep in memory (\cref{fig:input}, right) for the Controller using the metadata gathered from SQL executions in the DBMS and our proposed algorithm \optimizationproblem (described in \Cref{sec:problem_setup,sec:joint_optimization}).

\mypara{DBMS}

The DBMS executes SQL statements from Controller (e.g., joins, aggregation). Input tables are from either \memory (described shortly) or read from external storage (e.g., disk).
Our current implementation uses a Presto DBMS cluster~\cite{prestosql}, but any DBMS may be used in its place.

\subsection{Performing an MV Refresh Run}
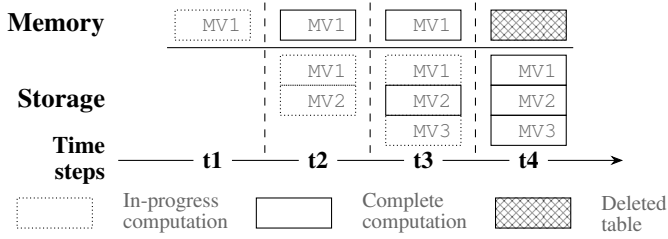
\begin{figure}[t]
\begin{subfigure}[b]{\linewidth}
\centering
\begin{tikzpicture}
 \node(mvdefinition)[minimum height=10mm, minimum width=31mm, anchor=south] at (10.8, 0){};

\begin{scope}[every node/.style={rectangle,thick, anchor=east}]
    \node[align=right] (label5) at (-0.7, 1) {\textbf{Memory}};
    \node[align=right] (label6) at (-0.7, 0) {\textbf{Storage}} ;
\end{scope}

\begin{scope}[every node/.style={rectangle, anchor=west, inner sep=0.05cm,fill opacity = 0.6,text opacity=1, minimum height = 4mm, minimum width = 10mm}]
    \node[densely dotted,draw=black] (X1) at (0.1, 1) {\footnotesize \texttt{MV1}};
    \node[draw=black] (X2) at ($(X1.west) + (1.4, 0)$) {\footnotesize \texttt{MV1}};
    \node[draw=black] (X3) at ($(X2.west) + (1.4, 0)$) {\footnotesize \texttt{MV1}};
    \node[draw=black,  pattern=crosshatch, pattern color = black] (X4) at ($(X3.west) + (1.4, 0)$) {};
    \node[densely dotted, draw=black] (Y1) at ($(X2.west) + (0, -0.6)$) {\footnotesize \texttt{MV1}};
    \node[densely dotted, draw=black] (Y2) at ($(Y1.west) + (1.4, 0)$) {\footnotesize \texttt{MV1}};
    \node[draw=black] (Y3) at ($(Y2.west) + (1.4, 0)$){\footnotesize \texttt{MV1}};
    \node[densely dotted,draw=black] (B1) at ($(Y1.west) + (0, -0.4)$) {\footnotesize \texttt{MV2}};
    \node[draw=black] (B2) at ($(B1.west) + (1.4, 0)$) {\footnotesize \texttt{MV2}};
    \node[draw=black] (B3) at ($(B2.west) + (1.4, 0)$) {\footnotesize \texttt{MV2}};
    \node[densely dotted,draw=black] (A1) at ($(B2.west) + (0, -0.4)$) {\footnotesize \texttt{MV3}};
    \node[draw=black] (A2) at ($(A1.west) + (1.4, 0)$) {\footnotesize \texttt{MV3}};

    \node[densely dotted,draw=black] (l1) at (-2.0, -1.5) {};
    \node[align=left] (l11) at ($(l1.east) + (0.1, 0)$) {\footnotesize In-progress\\[-0.4em]\footnotesize computation} ;
    \node[draw=black] (l2) at ($(l11.east) + (0.3, 0)$) {};
    \node[align=left] (l21) at ($(l2.east) + (0.1, 0)$) {\footnotesize Complete\\[-0.4em]\footnotesize computation};
    \node[fill=FlagColor,draw=black, pattern=crosshatch] (l3) at ($(l21.east) + (0.3, 0)$) {};
    \node[align=left] (l31) at ($(l3.east) + (0.1, 0)$) {\footnotesize Deleted \\[-0.4em] \footnotesize table};
\end{scope}

\node[align=right] (t1) at ($(X1) + (0, -1.8)$) {\small \textbf{t1}} ;
\node[align=right] (t2) at ($(t1) + (1.4, 0)$) {\small \textbf{t2}} ;
\node[align=right] (t3) at ($(t2) + (1.4, 0)$) {\small \textbf{t3}} ;
\node[align=right] (t4) at ($(t3) + (1.4, 0)$) {\small \textbf{t4}} ;
\node[align=right,anchor=east] (tlabel) at (-0.7, -0.8) {\small \textbf{Time}\\[-0.2em]\small \textbf{steps}} ;

\begin{scope}[>={Stealth[black]},
              every node/.style={fill=none,circle,dashed},
              every edge/.style={draw=black,dashed}]
    \draw[-,dashed] (1.3, 1.3) -- (1.3, -0.7);
    \draw[-,dashed] (2.7, 1.3) -- (2.7,-0.7);
    \draw[-,dashed] (4.1, 1.3) -- (4.1, -0.7);
\end{scope}

\begin{scope}[>={Stealth[black]},
              every node/.style={fill=none,circle},
              every edge/.style={draw=black}]
    \draw[-] (0, 0.7) -- (5.4, 0.7);
    \draw[-] ($(t1.west) + (-1.0, 0)$) -- ($(t1.west)$);
    \draw[-] ($(t1.east)$) -- ($(t2.west)$);
    \draw[-] ($(t2.east)$) -- ($(t3.west)$);
    \draw[-] ($(t3.east)$) -- ($(t4.west)$);
    \draw[->] ($(t4.east)$) -- ($(t4.east) + (1.0, 0)$);
\end{scope}

\end{tikzpicture}
\end{subfigure}
\vspace{-5mm}
\caption{The example MV refresh process by timestamp (t1-t4) for the workload given the plan in \cref{fig:input}.}
\label{fig:mv_refresh}
\vspace{-4mm}
\end{figure}
\label{sec:job_representation}
During an MV refresh run, Controller directs the DBMS to execute nodes (i.e. running their SQL statements) one by one according to the execution order computed by the Optimizer.

\mypara{Memory Management} 
\system ensures that the size of nodes kept in memory (called the \memory) is bounded correctly.
Let $n$ be a node that is to be kept in memory according to the Optimizer's plan: $n$ is created directly in the \memory, and freed from the \memory as soon as all the nodes depending on $n$ complete their execution.

\mypara{Parallelizing Compute and Materialization} 
\system aims to maximize the usage of compute and I/O bandwidth of a DBMS.
Nodes created in memory are materialized to external storage immediately after their creation;
this materialization process occurs in parallel with the execution of downstream nodes.
    

\mypara{Example}
Based on \cref{fig:input}, we present a concrete example in \cref{fig:mv_refresh} to illustrate how \system coordinates MV refreshes while maintaining the correct usage of the \memory and parallelizing I/O and compute.
The MVs are refreshed in the order of \texttt{MV1}, \texttt{MV2}, \texttt{MV3} according to the timeline below; \texttt{MV1} is the only MV to be kept in memory:\begin{itemize}
    \item \textbf{t1}: \texttt{MV1} is created in the \memory.
    \item \textbf{t2}: \texttt{MV1} is fully created in memory. It is materialized to disk concurrently while refreshing \texttt{MV2} (using the newly created \texttt{MV1}); \texttt{MV2} will be directly created on disk.
    \item \textbf{t3}: \texttt{MV2} is now fully refreshed. the refresh of \texttt{MV3} starts on storage reading from \texttt{MV1} in memory.
    \item \textbf{t4}: \texttt{MV3} is fully refreshed and \texttt{MV1} is fully materialized. \texttt{MV1} in memory is deleted, concluding the MV refresh run.
\end{itemize}



\section{\optimizationproblem: Problem Setup}
\label{sec:problem_setup}

This section formally defines \optimizationproblem, namely \textit{speeding up MV refresh runs via intermediate data storage under bounded memory}.
\Cref{tbl:symbols} lists symbols we use in describing \optimizationproblem.

\mypara{Problem Inputs} \optimizationproblem has 3 inputs:

\begin{enumerate}
    \item A directed acyclic graph (DAG) $\mathcal{G} := \{\mathcal{V}, \mathcal{E}\}$ of the MV refresh run.
    $\mathcal{V} = \{v_1,...,v_n\}$ is the set of $n$ individual MV updates (nodes), and $\mathcal{E}$ is the set of dependencies between the MV updates (edges). 
    \item The set of intermediate table sizes produced by each node $\mathcal{S} = \{s_1,...,s_n\}$, which we will abbreviate as \textit{node sizes}. 
    Each $s_i$ measures the amount of memory space required to keep the intermediate table produced by $v_i$.
    \item The set of speedup scores of each node $\mathcal{T} = \{t_1,...,t_n\}$, where $t_i$ measures the estimated speedup for the MV refresh run achieved by keeping the output of $v_i$ in memory (which we say \textit{flagging the node}).
\end{enumerate}

\begin{table}[t]
\centering
\caption{Table of Symbols}
\label{tbl:symbols}

\vspace{-2mm}
\small
\begin{tabular}{ll}
\toprule
\textbf{Symbols}              & \textbf{Definition}           \\ \midrule
$\mathcal{G} := \{\mathcal{V}, \mathcal{E}\}$         & The dependency graph                    \\
$\mathcal{V} = \{v_1,...,v_n\}$            & Set of $|\mathcal{V}| = n$ nodes       \\
$\mathcal{E}$            & Set of $|\mathcal{E}| = m$ dependencies            \\
\hline
$\mathcal{S} = \{s_1,...,s_n\}$        & Set of intermediate table (node) sizes \\
$\mathcal{T} = \{t_1,...,t_n\}$        & Set of speedup scores    \\\hline
$M$ & \memory size      \\\hline
$\tau: [1..n] \rightarrow [1..n]$ & MV refresh (execution) order   \\
$\mathcal{U} \subseteq \mathcal{V}$ & \makecell[l]{Nodes results kept in memory (flagged nodes)}     \\
\bottomrule
\end{tabular}
\vspace{-4mm}
\end{table}

\mypara{Speedup Scores} 
The speedup score $t_i$ of node $n_i$ is computed as follows:
\begin{multline*}
t_i = \Big( \sum\nolimits_{(v_i, v_j) \in \mathcal{E}} \text{data-access-time}(v_j | v_i\, \text{on disk}) 
    \\[-0.2em]-  \text{data-access-time}(v_j | v_i\, \text{in memory}) \Big) 
    \\[-0.2em] + \text{time}(\text{create}\, v_i\, \text{on disk}) 
        -  \text{time}(\text{create}\, v_i\, \text{in memory})
\end{multline*}

\noindent
Speedup $t_i$ of flagging $v_i$ is computed w.r.t.~to the baseline approach of performing MV refresh operations sequentially: 
For each downstream node $v_j$ depending on $v_i$, data access time is saved by reading $v_i$ from memory instead of from external storage. For executing $v_i$, write time can be reduced by materializing $v_i$ to external storage in parallel with minimal interference with downstream computations. (\Cref{sec:job_representation})

\mypara{Memory Usage and Execution Order}

Given the \memory size $M$, the peak memory usage of an MV refresh run---the maximum combined size of flagged nodes coexisting in memory---at any time during the run should not exceed $M$. 
The peak memory usage of a set of flagged nodes depends on the execution order $\tau: [1..n] \rightarrow [1..n]$ in which the nodes are executed: $v_i$ is the $\tau(i)^{th}$ executed node.

The execution order determines when a flagged node can be released from memory---a flagged node can be released only when all of its downstream nodes (children) are computed. 
Hence, it plays a key role in determining the feasibility of a set of flagged nodes under the \memory size constraint.  
A good execution order will allow us to release flagged nodes faster, which in turn allows us to flag more nodes throughout the MV refresh run, leading to a higher total speedup score under the same \memory size constraint.

\begin{figure}[t]
\begin{subfigure}[b]{\linewidth}
\centering
\begin{tikzpicture}
\begin{scope}[every node/.style={circle,thick,draw,align=center,inner sep=0.1mm}]
    \node[] (v1) at (-1, 0) {\footnotesize$v_1$\\[-0.3em]\footnotesize 100GB};
    \node[] (v2) at ($(v1) + (1.0, 0)$) {\scriptsize $v_2$\\[-0.3em]\scriptsize 10GB};
    \node[] (v3) at ($(v2) + (1.0, 0)$) {\footnotesize$v_3$\\[-0.3em]\footnotesize 100GB} ;
    \node[] (v4) at ($(v1) + (0.5, 1.0)$) {\scriptsize $v_4$\\[-0.3em]\scriptsize 10GB} ;
    \node[] (v5) at ($(v4) + (1.0, 0)$) {\scriptsize $v_5$\\[-0.3em]\scriptsize 10GB} ;
    \node[] (v6) at ($(v4) + (0.5, 1.0)$) {\scriptsize $v_6$\\[-0.3em]\scriptsize 10GB} ;
\end{scope}

\begin{scope}[>={Stealth[black]},
              every node/.style={fill=none,circle},
              every edge/.style={draw=black, thick}]
    \path [->] (v1) edge node {} (v4);
    \path [->] (v2) edge node {} (v4);
    \path [->] (v2) edge node {} (v5);
    \path [->] (v3) edge node {} (v5);
    \path [->] (v4) edge node {} (v6);
    \path [->] (v5) edge node {} (v6);
\end{scope}

\begin{scope}[>={Stealth[black]},
              every node/.style={fill=none,circle},
              every edge/.style={draw=ExampleColor2,line cap=round}]
    \node[] (r1) at (-1.7, -0.2) {};
    \node[] (r2) at (0.5, -0.2) {};
    \node[] (r3) at (0.5, 0.1) {};
    \node[] (r4) at (-0.1, 1.0) {};
    \node[] (r5) at (0, 1.2) {};
    \node[] (r6) at (0.1, 1.2) {};
    \node[] (r7) at (0.1, 1.4) {};
    \node[] (r8) at (-0.8, 2.6) {};
    \path [line width = 0.75mm] (r1.south) edge[] node {} (r2.south);
    \path [line width = 0.75mm] (r2.south) edge[bend right =60] node {} (r3.south);
    \path [line width = 0.75mm] (r3.south) edge[] node {} (r4.south);
    \path [line width = 0.75mm] (r4.south) edge[bend left = 60] node {} (r5.south);
    \path [line width = 0.75mm] (r5.south) edge[] node {} (r6.south);
    \path [line width = 0.75mm] (r6.south) edge[bend right = 60] node {} (r7.south);
    \path [line width = 0.75mm, arrows={->[ExampleColor2]}] (r7.south) edge node {} (r8.south);
\end{scope}
\node[text=red] (txt1) at (-0.9, 2.6) {\footnotesize Order 1 ($\tau_1$)} ;


\begin{scope}[>={Stealth[black]},
              every node/.style={fill=none,circle},
              every edge/.style={draw=ExampleColor1,line cap=round}]
    \node[] (b1) at (-1.7, 0.5) {};
    \node[] (b2) at (-0.3, 0.5) {};
    \node[] (b3) at (-0.2, 0.6) {};
    \node[] (b4) at (-0.4, 0.8) {};
    \node[] (b5) at (-0.3, 0.9) {};
    \node[] (b6) at (1.1, 0.5) {};
    \node[] (b7) at (1.2, 0.7) {};
    \node[] (b8) at (0.1, 2.6) {};
    \path [line width = 0.75mm] (b1.south) edge[] node {} (b2.south);
    \path [line width = 0.75mm] (b2.south) edge[bend right = 45] node {} (b3.south);
    \path [line width = 0.75mm] (b3.south) edge[] node {} (b4.south);
    \path [line width = 0.75mm] (b4.south) edge[bend left = 60] node {} (b5.south);
    \path [line width = 0.75mm] (b5.south) edge[] node {} (b6.south);
    \path [line width = 0.75mm] (b6.south) edge[bend right =60] node {} (b7.south);
    \path [line width = 0.75mm, arrows={->[ExampleColor1]}] (b7.south) edge[] node {} (b8.south);
\end{scope}
\node[text=blue] (txt2) at (0.7, 2.6) {\footnotesize Order 2 ($\tau_2$)} ;


\begin{scope}[every node/.style={rectangle,thick, anchor=east}]
    \node[align=right,text=red] (label1) at (6.5, 2.5) {\small Order 1};
    \node[align=right] (label2) at (4, 2.6) {\footnotesize Memory usage};
    \node[align=right] (label3) at (3.5, 2.3) {\footnotesize 100GB} ;
    \node[align=right,text=blue] (label4) at (6.5, 0.8) {\small Order 2};
    \node[align=right] (label5) at (4, 0.9) {\footnotesize Memory usage};
    \node[align=right] (label6) at (3.5, 0.6) {\footnotesize 100GB} ;
    \node[align=right] (label7) at (6.8, 1.3) {\footnotesize Node};
    \node[align=right] (label8) at (6.8, -0.4) {\footnotesize Node} ;
\end{scope}
    \node[text=blue] at (3.9, -0.4) {\small 1};
    \node[text=blue] at (4.3, -0.4) {\small 2};
    \node[text=blue] at (4.7, -0.4) {\small 4};
    \node[text=blue] at (5.1, -0.4) {\small 3};
    \node[text=blue] at (5.5, -0.4) {\small 5};
    \node[text=blue] at (5.9, -0.4) {\small 6};
    
    \node[text=red] at (3.9, 1.3) {\small 1};
    \node[text=red] at (4.3, 1.3) {\small 2};
    \node[text=red] at (4.7, 1.3) {\small 3};
    \node[text=red] at (5.1, 1.3) {\small 4};
    \node[text=red] at (5.5, 1.3) {\small 5};
    \node[text=red] at (5.9, 1.3) {\small 6};

\begin{scope}[>={Stealth[black]},
              every node/.style={fill=none,circle},
              every edge/.style={draw=black}]
    \draw[dashed] (3.5, 2.3) -- (6.5,2.3);
    \draw[dashed] (3.5, 0.6) -- (6.5,0.6);
    \draw[] (3.9, 1.5) -- (3.9, 1.45);
    \draw[] (4.3, 1.5) -- (4.3, 1.45);
    \draw[] (4.7, 1.5) -- (4.7, 1.45);
    \draw[] (5.1, 1.5) -- (5.1, 1.45);
    \draw[] (5.5, 1.5) -- (5.5, 1.45);
    \draw[] (5.9, 1.5) -- (5.9, 1.45);
    \draw[] (3.9, -0.2) -- (3.9, -0.25);
    \draw[] (4.3, -0.2) -- (4.3, -0.25);
    \draw[] (4.7, -0.2) -- (4.7, -0.25);
    \draw[] (5.1, -0.2) -- (5.1, -0.25);
    \draw[] (5.5, -0.2) -- (5.5, -0.25);
    \draw[] (5.9, -0.2) -- (5.9, -0.25);
    \draw[->] (3.5, 1.5) -- (3.5,2.5);
    \draw[->] (3.5, -0.2) -- (3.5,0.8);
    \draw[->] (3.5,1.5) -- (6.5, 1.5);
    \draw[->] (3.5,-0.2) -- (6.5, -0.2);
\end{scope}

    \node[draw,fill=ExampleColor2,anchor=south west,minimum width = 1.6cm, minimum height = 0.8cm,fill opacity = 0.6,text opacity=1] at (3.5, 1.5) {\large $v_1$};
    \node[draw,fill=ExampleColor2,inner sep=0.05cm,anchor=south west,minimum width = 0.8cm, minimum height = 0.1cm,fill opacity = 0.6,text opacity=1] at (5.1, 1.5) {\scriptsize $v_5$};
    \node[draw,fill=ExampleColor2,inner sep=0.05cm,anchor=south west,minimum width = 0.4cm, minimum height = 0.1cm,fill opacity = 0.6,text opacity=1] at (5.5, 1.75) {\scriptsize $v_6$};
    
    \node[draw,fill=ExampleColor1,anchor=south west,minimum width = 1.2cm, minimum height = 0.8cm,fill opacity = 0.6,text opacity=1] at (3.5, -0.2) {\large $v_1$};
    \node[draw,fill=ExampleColor1,anchor=south west,minimum width = 0.8cm, minimum height = 0.8cm,fill opacity = 0.6,text opacity=1] at (4.7, -0.2) {\large $v_3$};
    \node[draw,fill=ExampleColor1,inner sep=0.05cm,anchor=south west,minimum width = 0.4cm, minimum height = 0.1cm,fill opacity = 0.6,text opacity=1] at (5.5, -0.2) {\scriptsize $v_6$};

\end{tikzpicture}
\end{subfigure}
\vspace{-5mm}
\caption{Execution order matters: $v_1$ and $v_3$ cannot both be flagged under $\tau_1$, but can under $\tau_2$. The larger the combined size of flagged nodes, the better.}
\label{fig:toyexample}
\vspace{-4mm}
\end{figure}
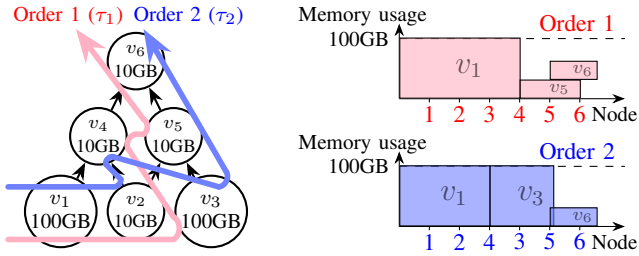

Consider the toy example shown in \cref{fig:toyexample}.
For simplicity, assume the speedup score of a node is equal to its size in GB; ideally, we want to flag both 100GB nodes $v_1$ and $v_3$ with \memory size $M=100GB$. 
As $v_1$ can be released after $v_4$ is executed, $\tau_2$ allows both $v_1$ and $v_3$ to be flagged by executing $v_4$ before $v_3$, while $\tau_1$ does not. The maximum possible  score of $210$ is achieved under $\tau_2$ by flagging both of $v_1$ and $v_3$ (and $v_6$). This is much higher than the maximum possible score of $120$ achieved by flagging $v_1, v_5, v_6$ under $\tau_1$.

\mypara{Problem Definition}

According to our problem setup, we define \optimizationproblem as a joint optimization problem of the set of flagged nodes $\mathcal{U}$ and execution order $\tau$:

\begin{problem}{\optimizationproblem}
\label{prof:optimization}
\begin{description}[leftmargin=1.25cm]
\item[\normalfont Input:\,\,\,\,\hspace{0.05em}]\begin{enumerate}
    \item Dependency graph $\mathcal{G} = \{\mathcal{V}, \mathcal{E}\}$
    \item Node sizes $\mathcal{S}$
    \item Speedup scores $\mathcal{T}$
    \item \memory size $M$
\end{enumerate}
\item[\normalfont Output:]\begin{enumerate}
    \item A subset $\mathcal{U} \subseteq \mathcal{V}$ of flagged nodes
    \item An execution order $\tau$
\end{enumerate}
\item[\normalfont Objective function:]
    Maximize total speedup score $\sum_{i: v_i \in \mathcal{U}} t_i$ of flagged nodes
\item[\normalfont Constraint:]
    Flagging $\mathcal{U}$ is feasible under execution order $\tau$ and \memory size $M$
\end{description}
\end{problem}

\section{\optimizationproblem: Joint Optimization
}
\label{sec:joint_optimization}

We present our solution to \optimizationproblem. 
Unlike previous work on 
scheduling refresh order of a set of MVs~\cite{folkert2005optimizing,golab2009scheduling} 
and storing select intermediate data to reduce job latency~\cite{xin2021enhancing, wangtempura, jalaparti2018netco}, 
we are, to the best of our knowledge, the first to 
consider joint optimization of both items.

\optimizationproblem is NP-hard via a reduction from the 0-1 knapsack problem, which necessitates the usage of an efficient approximation algorithm. While \optimizationproblem has 2 distinct outputs flagged nodes $U$ and execution order $\tau$, standard alternating optimization~\cite{bezdek2002some} is not applicable due to improving $\tau$ while holding $U$ constant does not improve the objective function of total speedup score. We still decompose \optimizationproblem into 2 distinct subproblems and solve \optimizationproblem by starting from an initial execution order $\tau_0$ 
and an empty set of flagged nodes (i.e., $\mathcal{U}_0 = \emptyset$); however, differing from standard alternating optimization, we decompose \optimizationproblem into 2 subproblems with different objective functions:

\begin{problem}{\subproblemnodes}
\label{prof:optimization_nodes}
\begin{description}
\item[\normalfont Input:] Inputs of \optimizationproblem, \textbf{Execution order $\tau$}
\item[\normalfont Output:] Flagged nodes $\mathcal{U}$
\item[\normalfont Objective function:] Maximize total speedup score $\sum_{v_i \in \mathcal{U}} t_i$
\item[\normalfont Constraint:] Same as \optimizationproblem
\end{description}
\end{problem}

\begin{problem}{\subproblemorder}
\label{prof:optimization_order}
\begin{description}
\item[\normalfont Input:] Inputs of \optimizationproblem, \textbf{Flagged nodes $\mathcal{U}$}
\item[\normalfont Output:] Execution order $\tau$
\item[\normalfont Objective function:] Minimize average memory usage $\frac{1}{n}\sum_{i:v_i\in\mathcal{U}}(max_{(v_i, v_j)\in \mathcal{E}}\tau(j) - \tau(i))s_i)$
\item[\normalfont Constraint:] Same as \optimizationproblem
\end{description}
\end{problem}

Instead of improving the objective function, we 'relax the constraints' in \subproblemorder by using the average memory usage as its objective function, which we define as the average size of flagged nodes stored in \memory assuming unit job execution times ( illustrated as shaded regions in \cref{fig:dfs}).
A lower average memory usage corresponds to more efficient usage of \memory through a more timely release of flagged nodes, potentially allowing us to add more nodes into $\mathcal{U}$ in future iterations of alternating optimization.

We optimize for average memory usage in \subproblemorder despite peak memory usage being the constraint of \optimizationproblem due to the latter failing to meaningfully differentiate between execution orders.
\cref{fig:dfs} showcases an example: nodes $v_1, v_3, v_4, v_5$ are flagged. 
While both refresh orders have the same peak memory usage, the refresh order found by our solution to \subproblemorder (\solutionorder, \cref{sec:optimize_order}) attains the peak memory for a shorter period of time - hence having a lower average memory usage - allowing an additional node to be flagged in the next iteration of solving \subproblemnodes.

In the following subsections, we describe our solutions to \subproblemnodes (\Cref{sec:optimize_nodes}) and \subproblemorder (\Cref{sec:optimize_order}).
Finally, we explain
how we combine them to form our alternating optimization algorithm for solving  \optimizationproblem (\Cref{sec:optimize_all}).

\subsection{\subproblemnodes: Solution}
\label{sec:optimize_nodes}

We solve \subproblemnodes exactly by formulating it as a multidimensional 0-1 knapsack problem (MKP)~\cite{kellerer2004multidimensional}. We have also considered approximate solutions such as greedy algorithms; however, they exhibit considerably worse performance compared to our MKP formulation while having a negligible runtime advantage. (\Cref{sec:exp_algm})

\mypara{MKP Setup} For each node $v_i$ and execution order $\tau$, let $\mathcal{V}_i$ denote a set of nodes that, when flagged, will be kept in memory during the time of $v_i$'s execution.
According to the memory management scheme of \system, a flagged node $v_j$ is kept in memory at the time of $v_i$'s execution if there is at least 1 child of $v_j$ yet to be executed: 
\[
\mathcal{V}_i := \{v_j| \tau(j) \leq \tau(i) \leq max_{(v_j, v_k) \in \mathcal{E}}\tau(k)\}
\]\\[-1.5em]
The set of all $\mathcal{V}_i$s form the constraints of the MKP such that the total size of flagged nodes in each $\mathcal{V}_i$ does not exceed \memory size $M$.
Using speedup scores $\mathcal{T}$ as coefficients of the 
 MKP's objective function, we define \subproblemnodes in integer linear programming format\footnote{We round speedup scores to the nearest integer.} as follows:
 \vspace{-1mm}
\begin{equation*}
\begin{array}{ll@{}ll}
\text{maximize}  & \displaystyle\sum\limits_{i=1}^{n} &x_{i}t_{i} &\\
\text{subject to}& \displaystyle\sum\limits_{j:v_{j} \in \mathcal{V}_{i}}   &x_{j}s_{j} \leq M,  &i=1 ,\dots, n\\[-0.3em]
                 &                                                &x_{i} \in \{0,1\}, &i=1 ,\dots, n
\end{array}
\end{equation*}

\newlength{\textfloatsepsave} 
\setlength{\textfloatsepsave}{\textfloatsep} \setlength{\textfloatsep}{1pt}
\begin{algorithm}[t]
\caption{SimplifiedMKP \label{lm:computeflaggednodes}}
\SetKwInOut{Input}{Input}
\SetKwInOut{Output}{Output}
\Input{
  (1) Dependency graph $\mathcal{G} = \{\mathcal{V}, \mathcal{E}\}$ \\
  (2) Node sizes $\mathcal{S}$ and speedup scores $\mathcal{T}$ \\
  (3) \memory size $M$ \\
  (4) Execution order $\tau$
}
\Output{Set of flagged nodes $\mathcal{U}$}
Initialize $\mathcal{V}_{exclude} = \{v_i | s_i > M \lor t_i = 0\}$; \\
\texttt{constraint\_sets}$ = \texttt{GetConstraints}(\mathcal{G}, \mathcal{S}, \mathcal{V}_{exclude}, M, \tau)$; \\
Initialize $\mathcal{V}_{mkp} = \bigcup_{\mathcal{V}_i \in \texttt{constraint\_sets}}\mathcal{V}_i$; \\
Initialize $|$\texttt{constraint\_sets}$| = k, |\mathcal{V}_{mkp}| = l$;\\
Set profits $P_{1\times l} = [t_{i_1},...,t_{i_l}], v_{i_1},...,v_{i_l} \in \mathcal{V}_{mkp}$;\\
Set weights $W_{k\times l}, w_{xy} = (\text{if } v_{i_y} \in \mathcal{V}_x \text{ then } s_i \text{ else } 0),1\leq x\leq k, 1\leq y \leq l$;\\
Set capacities $C_{k\times 1} = [M,...,M]$;\\
$\mathcal{U} = \texttt{BinaryMKPSolver}(P, W, C)$;\\
$\mathcal{U} \leftarrow \mathcal{U} \cup (\mathcal{V} - \mathcal{V}_{mkp} - \mathcal{V}_{exclude})$;\\
\textbf{Return} the set of nodes to store in memory $\mathcal{U}$.
\end{algorithm}

\mypara{Simplifying the MKP} We proceed to simplify the MKP by identifying and removing redundant variables (nodes) present in the MKP. A node $v_i$ is redundant and does not need to be evaluated in the MKP if either:\begin{itemize}
    \item $s_i > M$: $v_i$ has size larger than \memory size ---flagging $v_i$ by itself violates the constraint
    \item $t_i = 0$: storing $v_i$ does not contribute to the objective
\end{itemize}
We can exclude such nodes from our constraint sets $\mathcal{V}_i$:
\begin{multline*}
\mathcal{V}_i := \{v_j| \tau(j) \leq \tau(i) \leq max_{(v_j, v_k) \in \mathcal{E}}\tau(k) \land v_j \not \in \mathcal{V}_{exclude}\}\\[-0.2em]
\text{where } \mathcal{V}_{exclude} := \{v_i |i: s_i > M \lor t_i = 0 \}
\end{multline*}
A constraint set $\mathcal{V}_i$ is redundant and does not need to be evaluated in the MKP if either:\begin{itemize}
    \item $\mathcal{V}_i$ is not maximal: $\exists~ \mathcal{V}_j s.t. \mathcal{V}_i \subsetneq \mathcal{V}_j$; the constraint set $\mathcal{V}_i$ is a strict subset of another constraint set $\mathcal{V}_j$
    \item $\mathcal{V}_i$ is trivial: $\sum_{j:v_j\in \mathcal{V}_i} s_j \leq M$; the constraint $\mathcal{V}_i$ cannot be violated even if all nodes in $\mathcal{V}_i$ are flagged
\end{itemize}

\ignore{
\paragraph{Example MKP simplification} Consider again the toy example in \cref{fig:toyexample}. Under execution order $\tau_2$ and \memory size $M = 100GB$, the constraint sets are as follows:
\begin{multicols}{2}
\begin{itemize}
    \item $\mathcal{V}_1 := \{v_1\}$
    \item $\mathcal{V}_2 := \{v_1, v_2\}$
    \item $\mathcal{V}_4 := \{v_1, v_2, v_4\}$
    \item $\mathcal{V}_3 := \{v_2, v_3, v_4\}$
    \item $\mathcal{V}_5 := \{v_2, v_3, v_4, v_5\}$
    \item $\mathcal{V}_6 := \{v_4, v_5, v_6\}$
\end{itemize}
\end{multicols}
Out of the constraint sets, $\mathcal{V}_1, \mathcal{V}_2, \mathcal{V}_3$ are not maximal, $\mathcal{V}_1, \mathcal{V}_2, \mathcal{V}_6$ are trivial, hence $\mathcal{V}_4$ and $\mathcal{V}_5$ are the only non-redundant constraints.
}

\mypara{Solving the MKP} 

We present our completed solution (\solutionnodes) to \subproblemnodes in \cref{lm:computeflaggednodes}.
The set of all relevant (maximal and non-trivial) constraint sets is efficiently computed with \texttt{GetConstraints} in linear time by a linear scan over the nodes (line 2).
Note that not every non-excluded node $v_i \in \mathcal{V} - \mathcal{V}_{exclude}$ appears in a constraint set. If a node does not appear in any constraint set and is also not in $\mathcal{V}_{exclude}$, it can be trivially flagged 
because doing so will not violate memory constraints; 
these nodes are manually added to the solution returned from the MKP solver (line 9).

We choose branch-and-bound method\footnote{Our \system implementation uses the BnB solver from Google OR-Tools~\cite{ortools}.} for subroutine \texttt{BinaryMKPSolver} (line 8). While the method has worst-case exponential time complexity, we empirically verify that \cref{lm:computeflaggednodes} has satisfactory scalability in general (\Cref{sec:exp_complexity}): it scales roughly linearly with the number of nodes in the dependency graph and takes on average 0.02 seconds to find the optimal set of nodes to flag in a 100-node graph.
\subsection{\subproblemorder: Solution}
\label{sec:optimize_order}


We solve \subproblemorder by formulating a \textit{memory-aware} DFS-based scheduling algorithm. 
We have experimented with other methods, such as hill-climbing~\cite{seitz2010contributions} and recursive graph separators~\cite{charikar2010l}\footnote{While there exists an ILP formulation for this problem, ~\cite{baruah2022ilp}, it contains $O(n^3)$ variables and constraints making it too inefficient for real-time scheduling. We defer exploration of this direction to future work.}, though we find them to be less effective in comparison. (\Cref{sec:exp_algm})
\mypara{Optimizing Average Memory Usage} 
The DFS-based scheduling algorithm aims to minimize the time between a node's execution and its children's by finishing a branch of execution before starting a new one, which 
minimizes average memory usage of flagged nodes by freeing them as soon as possible.
 \setlength{\textfloatsep}{\textfloatsepsave}
\subfile{plots/fig_dfs}
However, we find that off-the-shelf DFS-based sorts in existing work~\cite{marchal2018parallel} are insufficient in our case. 
In particular, DFS traversals must choose a path to proceed onto (tie-break) at branches, and a random selection without considering the node size can result in suboptimal performance. 
A random selection may keep large flagged nodes in memory for an extended period of time, which reduces the amount of memory space available for flagging additional nodes (see \cref{fig:dfs}). 

\mypara{Memory-Aware DFS} To address this, we propose a \textit{memory-aware} DFS-based scheduling algorithm (\solutionorder),
    which prioritizes nodes with lower actual memory consumption to tie-break:
A node's actual memory consumption is equal to its size if it is flagged and is equal to 0 otherwise.
\solutionorder aims to compute the largest flagged dependencies (in terms of size) of a node last in order to minimize the time these large dependencies are kept in memory, which frees up memory space to potentially flag more nodes.

The toy example in \cref{fig:dfs} illustrates this idea.
We have \memory size $M=100GB$; for simplicity, assume the speedup score of a node is equal to its size in GB.
When performing tie-breaking between $v_2$ and $v_3$, \solutionorder schedules $v_2$ first as $v_3$ has higher actual memory consumption; despite being larger than $v_3$, $v_2$ is not flagged.
$v_3$ is kept in memory for only 3 node executions ($v_3$, $v_5$, $v_6$), freeing memory space for $v_6$ to be additionally flagged.
This would not be possible with a DFS-based sort with random tie-breaking in cases where it schedules $v_3$ instead of $v_2$ first. Such a sort would keep $v_3$ in memory for 5 node executions, taking up memory space that could have been used for flagging $v_6$.


We verify the validity of \solutionorder's execution order in each iteration of alternating optimization to preserve the memory constraint $M$.
In a rare case where \solutionorder outputs an infeasible execution order---violating the memory constraint $M$---we conclude that the execution order from the previous iteration of alternating optimization is optimal and terminate the alternating optimization algorithm.
While our solution may find locally optimal solutions, we empirically verify that the local optimums that our solution finds are still of high quality compared to solutions found by other methods. (\cref{sec:exp_algm})

\subsection{\optimizationproblem: Putting Everything Together}
\label{sec:optimize_all}

With solutions to both subproblems, we present our alternating optimization algorithm for solving \optimizationproblem in \cref{lm:jointoptimization}.

\setlength{\textfloatsep}{1pt}
\begin{algorithm}[t]
\caption{AlternatingOptimization} 
\label{lm:jointoptimization}
\SetKwInOut{Input}{Input}
\SetKwInOut{Output}{Output}
\Input{(1) Dependency graph $\mathcal{G} = \{\mathcal{V}, \mathcal{E}\}$\\
(2) node sizes $\mathcal{S}$ and speedup scores $\mathcal{T}$\\
(3) \memory size $M$}
\Output{(1) Set of flagged nodes $\mathcal{U}$\\
(2) execution order $\tau$}
Initialize $\tau$ = $\texttt{GetTopologicalOrder(G)}$;\\
Initialize $\mathcal{U} = \emptyset$;\\
\While{the algorithm does not converge}{
    $\mathcal{U}_{new} = \texttt{SimplifiedMKP}(\mathcal{G}, \mathcal{S}, \mathcal{T}, M, \tau)$;\\
    \lIf{$\sum_{i:v_i\in \mathcal{U}_{new}} s_i \leq \sum_{i:v_i\in \mathcal{U}} s_i$}{
        \!\!\! \textbf{return} $\mathcal{U}, \tau$
    }
    $\mathcal{U} \leftarrow \mathcal{U}_{new}$;\\
    $\tau_{new} = \texttt{MA-DFS}(\mathcal{G}, \mathcal{S}, \mathcal{T}, M, \mathcal{U})$\\
    \lIf{$\texttt{PeakMemoryUsage}(\mathcal{G}, \mathcal{S}, \mathcal{U}, \tau_{new}) > M$}{
        \!\!\! \textbf{return} $\mathcal{U}, \tau$
    }
    $\tau \leftarrow \tau_{new}$;\\
}
\end{algorithm}

Any topological sorting algorithm can be used as the \texttt{GetTopologicalOrder} subroutine for obtaining the initial execution order on line 1.\footnote{Our implementation of \system uses the $\texttt{topological\_sort}$ function in the Python NetworkX package~\cite{networkx}.}
In each iteration of alternating optimization, \texttt{PeakMemoryUsage} on line 8 is efficiently computed in linear time by a linear scan over the nodes.
We alternate between solving \subproblemnodes and \subproblemorder we can no longer improve our outputs: \begin{itemize}
    \item No better set of flagged nodes $\mathcal{U}_{new}$ is found by \texttt{SimplifiedMKP} (line 5)
    \item The execution order $\tau_{new}$ found by \texttt{MA-DFS} violates the memory constraint $M$ (line 8)
\end{itemize}
Our solution is guaranteed to converge as the total speedup score of $\mathcal{U}$ must increase in each iteration for alternating optimization to continue. 
Our solution typically converges in $<$10 iterations for dependency graphs with up to 100 nodes. 
 \setlength{\textfloatsep}{\textfloatsepsave}

\section{Experiments}
\label{sec:experiments}



In this section, we empirically study the effectiveness of \system. 
We make the following claims: 
\begin{enumerate}
\item \textbf{Faster end-to-end MV refresh:} \system is capable of speeding up \emph{end-to-end MV refresh time} by 1.04$\times$--5.08$\times$ compared to an unoptimized raw Presto DBMS, which is up to an additional 2.22$\times$ compared to other existing off-the-shelf methods. (\Cref{sec:exp_overall}) 
\item \textbf{Consistent execution time savings:} \system's optimization results in consistent savings across different dataset sizes ranging from 10GB up to 1TB. (\Cref{sec:exp_size})
\item \textbf{Effective utilization of memory: } \system effectively utilizes varying \memory sizes. \system's \memory can be allocated from spare memory in the system or from query memory with minimal loss of speedup. (\Cref{sec:exp_memory})
\item \textbf{Table read time reduction:} \system's storage of intermediate tables in memory results in a $1.42\times-1.51\times$ reduction in the table read latency. (\Cref{sec:exp_io})
\item \textbf{Effectiveness of solution to \optimizationproblem:} Solving \optimizationproblem with our solution (\solutionnodes + \solutionorder) results in a 1.06$\times$--1.23$\times$ total time reduction compared to ablated methods. (\Cref{sec:exp_algm})
\item \textbf{Scalability of \system on distributed setups}: \system is capable of achieving consistent speedup on multi-node DB clusters regardless of the number of nodes. (\Cref{sec:exp_distributed})
\item \textbf{Scalability of \system on complex workloads}: Our solution to \optimizationproblem has a negligible $<0.02s$ optimization time on complex DAGs with up to 100 nodes. (\Cref{sec:exp_complexity})
\end{enumerate}

\subsection{Experiment Setup}
\label{sec:exp:setup}
\begin{figure*}[t]
\usetikzlibrary{patterns}
\begin{subfigure}[b]{0.48\linewidth}
\begin{tikzpicture}

\pgfplotstableread[col sep=comma,]{
name
I/O 1
I/O 2
I/O 3
Cmpt 1
Cmpt 2
}\datatable

\begin{axis}[
    ybar stacked,
    clip=false,
    legend style={ 
        legend columns = 7,
        at={(-0.1,1.5)},
        anchor=south west,
        column sep=0cm,
    },  
    xtick={4, 9, 14, 19.25, 24.25},
    xticklabels from table={\datatable}{name},
                 x tick label style={anchor=east,xshift=2ex, yshift = -1ex},
    xlabel style={yshift = 1ex},
    width=90mm,
    height=35mm,
    bar width=1.2mm,
    ymin=0,
    axis y line*=none,
    axis x line*=none,
    ytick={150, 300, 450, 600},
    yticklabels={150, 300, 450, 600},
    ymax = 600,
    xmin = 0,
    xmax = 27,
    ymajorgrids,
    tick label style={font=\footnotesize},
    legend style={
        font=\footnotesize,
        /tikz/every even column/.append style={column sep=0.5cm},
        at={(0, 1.1)}, anchor=south west
    },
    label style={font=\footnotesize},
    xlabel={Workload},
    ylabel={Time (s)},
    area legend
    ]
                 
    
    \addplot
    [NoOptColor,fill=NoOptColor,x tick label style={xshift=-0.3cm}] table[x=Method,y=M1] {sections/plots/vs_established_methods1.txt};
    \addlegendentry[]{No optimization};
    
    \addplot [LRUColor,fill=LRUColor,x tick label style={xshift=-0.3cm}] table[x=Method,y=M2] {sections/plots/vs_established_methods1.txt};
    \addlegendentry[]{LRU Cache};
    
    \addplot [RandomColor,fill=RandomColor,x tick label style={xshift=-0.3cm}] table[x=Method,y=M3] {sections/plots/vs_established_methods1.txt};
    \addlegendentry[]{Random};
    
    \addplot
    [GreedyColor,fill=GreedyColor,x tick label style={xshift=-0.3cm}] table[x=Method,y=M4] {sections/plots/vs_established_methods1.txt};
    \addlegendentry[]{Greedy};
    
    \addplot [HeuristicColor,fill=HeuristicColor,x tick label style={xshift=-0.3cm}] table[x=Method,y=M5] {sections/plots/vs_established_methods1.txt};
    \addlegendentry[]{Ratio-based selection};
    
        \addplot
    [black,fill=SCColor,x tick label style={xshift=-0.3cm}] table[x=Method,y=M6] {sections/plots/vs_established_methods1.txt};
    \addlegendentry[]{\textbf{\system (Ours)}};

\end{axis}
    
\end{tikzpicture}

\vspace{-2mm}
\caption{100GB \tpcds, 1.6GB \memory}

\end{subfigure}
\begin{subfigure}[b]{0.48\linewidth}
\begin{tikzpicture}

\pgfplotstableread[col sep=comma,]{
name
I/O 1
I/O 2
I/O 3
Cmpt 1
Cmpt 2
}\datatable

\begin{axis}[
    ybar stacked,
    clip=false,
    xtick={4, 9, 14, 19.25, 24.25},
    xticklabels from table={\datatable}{name},
                 x tick label style={anchor=east,xshift=2ex, yshift = -1ex},
    xlabel style={yshift = 1ex},
    width=90mm,
    height=35mm,
    bar width=1.2mm,
    ymin=0,
    axis y line*=none,
    axis x line*=none,
    ytick={75, 150, 225, 300},
    yticklabels={75, 150, 225, 300},
    ymajorgrids,
    ymax = 300,
    xmin = 0,
    xmax = 27,
    tick label style={font=\footnotesize},
    label style={font=\footnotesize},
    xlabel={Workload},
    ylabel={Time (s)},
    area legend
    ]
                 
    
    \addplot
    [NoOptColor,fill=NoOptColor,x tick label style={xshift=-0.3cm}] table[x=Method,y=M1] {sections/plots/vs_established_methods2.txt};

    \addplot [LRUColor,fill=LRUColor,x tick label style={xshift=-0.3cm}] table[x=Method,y=M2] {sections/plots/vs_established_methods2.txt};

    \addplot [RandomColor,fill=RandomColor,x tick label style={xshift=-0.3cm}] table[x=Method,y=M3] {sections/plots/vs_established_methods2.txt};

    \addplot
    [GreedyColor,fill=GreedyColor,x tick label style={xshift=-0.3cm}] table[x=Method,y=M4] {sections/plots/vs_established_methods2.txt};

    \addplot [HeuristicColor,fill=HeuristicColor,x tick label style={xshift=-0.3cm}] table[x=Method,y=M5] {sections/plots/vs_established_methods2.txt};
    
            \addplot
    [black,fill=SCColor,x tick label style={xshift=-0.3cm}] table[x=Method,y=M6] {sections/plots/vs_established_methods2.txt};

\end{axis}
    
\end{tikzpicture}

\vspace{-2mm}
\caption{100GB \tpcdsdate, 0.8GB \memory}

\end{subfigure}

\caption{
End-to-end MV refresh times on 100GB datasets. \system achieves 1.04x--5.08x speedup with 1.6/0.8GB \memory compared to an unoptimized raw engine.
This is an additional up to 2.22x speedup compared to other off-the-shelf methods. 
}
\label{fig:vs_established_methods}
\end{figure*}
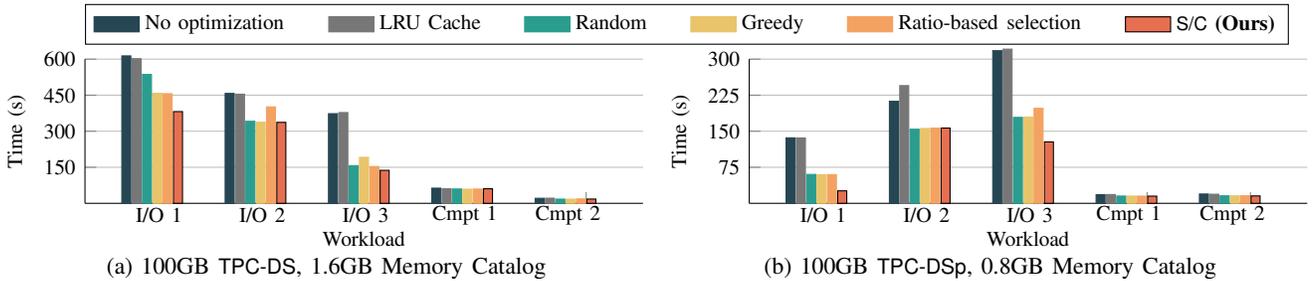
\mypara{Dataset} 

We use the table generator and queries included in the \textit{TPC-DS}~\cite{tpcds} decision support benchmark in our experiments.
We generate TPC-DS datasets from 5 distinct scale factors (10, 25, 50, 100, 1000); the scale factor determines the total size in GB of the tables in the generated dataset.

We create two copies of each dataset for each scale. 
One is a normal dataset generated as is (i.e., \tpcds). 
The other is a \textit{date-partitioned} dataset wherein
the three largest tables (\texttt{store\_sales, catalog\_sales, web\_sales}) are partitioned by year based on a join with the \texttt{date\_dim} table (i.e., \tpcdsdate).

\mypara{Methods} 

We evaluate the effectiveness of the \optimizationproblem framework by comparing it to both off-the-shelf algorithms and previously proposed algorithms:
\begin{itemize}
    \item \greedy: Iterate through nodes in execution order and flag nodes if doing so doesn't violate the memory constraint.
    \item \random: Iterate through nodes in random order and flag nodes if doing so doesn't violate the memory constraint.
    \item \ratio~\cite{xin2021enhancing}: Heuristic method prioritizing flagging nodes with high speedup score/node size ratio.
    \item The LRU cache in the DBMS caches query results. We increase the size of the LRU cache by an amount equal to the size of \memory.
\end{itemize}

\begin{table}[h]
\caption{Summary of workloads. I/O percentage estimated from profiling equivalent operations with Python Polars~\cite{polars}.}
\centering
\begin{tabular}{llll}
\toprule
Workload  & TPC-DS Queries & \# Nodes & I/O ratio \\
\midrule
I/O 1     & 5, 77, 80 & 21       & 51.5\%              \\
I/O 2     & 2, 59, 74, 75 & 19       & 59.0\%              \\
I/O 3     & 44, 49 & 26       & 46.6\%              \\
\hline
Compute 1 & 33, 56, 60, 61 & 21       & 0.9\%               \\
Compute 2 & 14, 23 & 16       & 28.3\%             \\
\bottomrule
\end{tabular}
\vspace{-4mm}
\label{tbl:workload}
\end{table}

We additionally perform an ablation analysis to evaluate the effectiveness of our solution to \system. 
Specifically, we evaluate our individual solutions to the subproblems \subproblemnodes and \subproblemorder by swapping out one subproblem solution for a baseline method during alternating optimization.

We compare our \solutionnodes solution to \subproblemnodes to the \greedy and \random methods.
We compare our \solutionorder solution to \subproblemorder to two baseline methods:
\begin{itemize}
    \item Simulated annealing (\simulatedannealing)~\cite{seitz2010contributions}:
    A hill-climbing algorithm that iteratively improves an execution order. In each iteration, two swappable nodes (i.e. doing so doesn't violate dependencies) are randomly selected; A swap is performed if doing so decreases the average memory usage. The swap is still performed with a certain probability to escape possible local minima. We set the iteration count to 10,000. 
    \item \separator~\cite{ravi1991ordering, rao2005new}:
    A divide-and-conquer algorithm that recursively finds separators/cuts in the DAG to partition nodes. In each iteration, a subgraph is partitioned into two via a cut; the algorithm stops when the original DAG has been partitioned into a series of singleton nodes by the cuts. These cuts define the execution order.
    This algorithm notably offers a lower bound of $O(n \log(n))$.
\end{itemize}


\mypara{Real Workload} We construct MV refresh workloads
from TPC-DS queries. 
We create one node/MV for each \textit{select-project-join} (SPJ) unit in the query and merge graphs of TPC-DS queries that share similar intermediate nodes and topics (i.e. profit report, sales analysis) into one larger graph.
Each graph represents a set of MVs to refresh together in a workload.
As a result, we obtain five workloads with dependency graphs, each consisting of 16--26 nodes (\Cref{tbl:workload}).

\mypara{Generated Workload} To investigate the scalability of our solution on more complex workloads, we create a workload generator for systematically creating realistic dependency graphs with a larger number of nodes (25--100).

Our workload generator consists of two components:\begin{itemize}
    \item A DAG generator for determining dependency between individual jobs in the workload~\cite{6471969}. We set generation parameters based on empirical analysis of real-world query structures (TPC-DS and Spider~\cite{yu-etal-2018-spider}).
    \item A Markov chain---trained on the same query set---for determining node operations (i.e. JOIN, AGG).
    Operations are used to derive the sizes and speedup scores of nodes from their inputs.
    The sizes of nodes with no parents (i.e. reading directly from base tables) are randomly sampled from table sizes in the 100GB \tpcds dataset.
\end{itemize}

\mypara{Environment} 

All experiments are performed on an Ubuntu server with an AMD EPYC 7552 48-Core Processor and 1TB RAM. The disk read and write speeds are 519.8 MB/s and 358.9 MB/s, respectively,
with a read latency of 175 $\mu s$. For most of our experiments, the Presto cluster consists of 1 worker node with 50GB/300GB query memory for the $<$100GB/1TB datasets respectively; however, we also study with distributed setups in \cref{sec:exp_distributed}. \system's \memory is independent of query memory unless otherwise stated.
%
\mypara{Implementation} We implement \system as a Python front-end over a Presto DBMS. The front-end performs optimization and sends queries to the DBMS via the Presto python connector~\cite{prestopython}; the Presto DBMS is created on top of a Hive Metastore~\cite{hive}, which stores created tables in an NFS.

We control data placement in the DBMS by creating tables either (i) in the Hive catalog~\cite{prestohive} to materialize or (ii) in the memory catalog ~\cite{prestomemory} to keep in memory. Base TPC-DS tables are stored in the ORC format~\cite{orc} native to Apache Hive. We create MVs/intermediate tables in the Apache Parquet format.

\mypara{Time Measurement} We measure the end-to-end time as the time from starting the MV refresh workload to having all MVs in the workload materialized on our NFS. 
For each setting, we measure the average end-to-end runtime over 5 runs
to reduce variance and report the median. 

The solution runtime for solving \optimizationproblem is not included in the end-to-end time as (i) optimization can be performed prior to data ingestion and (ii) solution runtimes are negligibly short compared to end-to-end MV refresh times. 

\subsection{Faster End-to-End MV Refresh}
\label{sec:exp_overall}

In this section, we study the impact of \system's optimization on
end-to-end MV refresh times.
For this,
we compare the average runtimes of our constructed TPC-DS workloads using Presto DBMS with \system, without \system, and with other algorithms on the 100GB TPC-DS dataset.

We report the results in \cref{fig:vs_established_methods}. The design goal of \system is to reduce the overhead of intermediate table read/write times, which is successfully reflected in our speedups achieved on the I/O heavy workloads, as follows. 
Compared to executing the workloads in serial fashion without keeping any data in memory, \system is able to speed up end-to-end workload execution time by 1.05$\times$--2.72$\times$ using 1.6GB \memory.
The savings increase to up to 1.20$\times$--5.08$\times$ on datasets with smaller intermediate table sizes (100GB date-partitioned TPC-DS datasets), which \system takes full advantage of to keep more data in memory throughout workload executions.

Compared to other algorithms, \system achieves up to 2.22$\times$ speedup in end-to-end workload execution time. Specifically, the performance of \system over other methods can be attributed to (i) \system additionally optimizing the execution order of the workload and (ii) greedy/heuristic selection can perform arbitrarily badly on workloads.
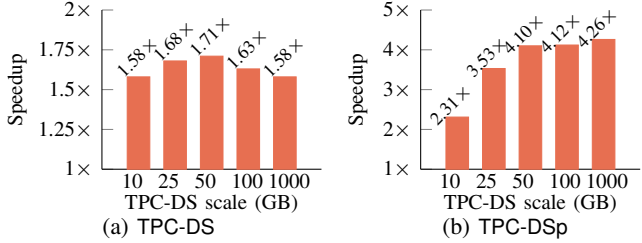
\begin{figure}[t]
\vspace{-2mm}
\begin{subfigure}[b]{0.48\linewidth}

\begin{tikzpicture}
\definecolor{TraditionalColor}{HTML}{ABDDA4}
\definecolor{MemcontainerColor}{HTML}{38A528}

\pgfplotstableread[]{
name
10
25
50
100
1000
}\datatable

\begin{axis}[
    ybar stacked,
                     legend style={ legend columns=4,
                                at={(-0.1,1.3)},anchor=south west,
                                column sep=0.1cm},  
    xtick=data,
    width=45mm,
    height=37mm,
    bar width=3mm,
    ymin=0,
    axis y line*=none,
    axis x line*=none,
    xticklabels from table={\datatable}{name},
    x tick label style={anchor=west,xshift=-2.2ex, yshift = -1ex},
    ytick={1.0, 1.25, 1.5, 1.75, 2},
    yticklabels={1$\times$, 1.25$\times$, 1.5$\times$, 1.75$\times$, 2$\times$},
    ymax = 2,
    ymin = 1,
    xmax = 6,
    xmin = 0,
    tick label style={font=\footnotesize},
    legend style={font=\footnotesize,yshift=-6ex},
    label style={font=\footnotesize},
    ylabel={Speedup},
    y label style={
        at={(axis description cs:-0.3,.5)},anchor=south
    },
    xlabel={TPC-DS scale (GB)},
    xlabel style={yshift = 1ex},
    area legend
    ]    
    
    \node[anchor=south west, rotate = 45] at (axis cs: 0.65, 1.48) {\scriptsize 1.58$\times$};
    \node[anchor=south west, rotate = 45] at (axis cs: 1.65, 1.58) {\scriptsize 1.68$\times$};
    \node[anchor=south west, rotate = 45] at (axis cs: 2.65, 1.61) {\scriptsize 1.71$\times$};
    \node[anchor=south west, rotate = 45] at (axis cs: 3.65, 1.53) {\scriptsize 1.63$\times$};
    \node[anchor=south west, rotate = 45] at (axis cs: 4.65, 1.48) {\scriptsize 1.58$\times$};
    
    \addplot [SCColor,fill=SCColor] table[x=Setting,y=speedup] {sections/plots/scale_tpcds.txt};

    \end{axis}
\end{tikzpicture}

\vspace{-3mm}
\caption{\tpcds}
\label{fig:scale_sweep_a}
\end{subfigure}
\hfill
\begin{subfigure}[b]{0.48\linewidth}
\hspace{-2mm}
\begin{tikzpicture}
\definecolor{TraditionalColor}{HTML}{ABDDA4}
\definecolor{MemcontainerColor}{HTML}{38A528}

\pgfplotstableread[col sep=comma,]{
name
10
25
50
100
 1000
}\datatable

\begin{axis}[
    ybar stacked,
                     legend style={ legend columns=4,
                                at={(-0.1,1.3)},anchor=south west,
                                column sep=0.1cm},  
    xtick=data,
    width=45mm,
    height=37mm,
    bar width=3mm,
    ymin=0,
    axis y line*=none,
    axis x line*=none,
    xticklabels from table={\datatable}{name},
    x tick label style={anchor=west,xshift=-2.28ex, yshift = -1ex},
    ytick={1, 2, 3, 4, 5},
    yticklabels={1$\times$, 2$\times$, 3$\times$, 4$\times$, 5$\times$},
    ymax = 5,
    ymin = 1,
    xmax = 6,
    xmin = 0,
    tick label style={font=\footnotesize},
    legend style={font=\footnotesize,yshift=-6ex},
    label style={font=\footnotesize},
    ylabel={Speedup},
        y label style={at={(axis description cs:-0.2,.5)},anchor=south},
    xlabel={TPC-DS scale (GB)},
    xlabel style={yshift = 1ex},
    area legend
    ]    
    
    \node[anchor=south west, rotate = 45] at (axis cs: 0.5, 1.71) {\scriptsize 2.31$\times$};
    \node[anchor=south west, rotate = 45] at (axis cs: 1.5, 2.93) {\scriptsize 3.53$\times$};
    \node[anchor=south west, rotate = 45] at (axis cs: 2.5, 3.50) {\scriptsize 4.10$\times$};
    \node[anchor=south west, rotate = 45] at (axis cs: 3.5, 3.52) {\scriptsize 4.12$\times$};
    \node[anchor=south west, rotate = 45] at (axis cs: 4.5, 3.66) {\scriptsize 4.26$\times$};
    
    \addplot [SCColor,fill=SCColor] table[x=Setting,y=speedup] {sections/plots/scale_tpcds_date.txt};

    \end{axis}
\end{tikzpicture}

\vspace{-3mm}
\caption{\tpcdsdate}
\label{fig:scale_sweep_b}
\end{subfigure}
\caption{\system's optimization achieves consistent speedup on a range of dataset scales. \memory set to 1.6\% data size.
}
\vspace{-4mm}
\label{fig:scale_sweep}
\end{figure}
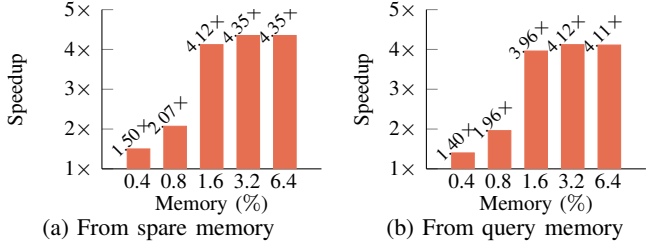
\begin{figure}[t]
\begin{subfigure}[b]{0.48\linewidth}
\begin{tikzpicture}
\definecolor{TraditionalColor}{HTML}{ABDDA4}
\definecolor{MemcontainerColor}{HTML}{38A528}

\pgfplotstableread{
name
0.4
0.8
1.6
3.2
6.4
}\datatable

\pgfplotstableread{
key val
1 1.50
2 2.07
3 4.12
4 4.35
5 4.35
}\numbertable

\begin{axis}[
    ybar stacked,
                     legend style={ legend columns=2,
                                at={(-0.3,1.65)},anchor=south west,
                                column sep=0.1cm},  
    xtick=data,
    width=45mm,
    height=37mm,
    bar width=3mm,
    ymin=0,
    axis y line*=none,
    axis x line*=none,
    xtick={1, 2, 3, 4, 5},
    xticklabels from table={\datatable}{name},
    x tick label style={anchor=east,xshift=1.8ex, yshift = -1ex},
    ytick={1, 2, 3, 4, 5},
    yticklabels={1$\times$, 2$\times$, 3$\times$, 4$\times$, 5$\times$},
    ymax = 5,
    ymin = 1,
    xmax = 6,
    xmin = 0,
    tick label style={font=\footnotesize},
    legend style={font=\footnotesize,yshift=-6ex},
    label style={font=\footnotesize},
    ylabel={Speedup},
    y label style={at={(axis description cs:-0.3,.5)},anchor=south},
    xlabel={Memory ($\%$)},
    xlabel style={yshift = 1ex},
    area legend
    ]    
    
    
    \addplot [SCColor,fill=SCColor] table[x=key,y=val] {\numbertable};

    \node[anchor=south west, rotate = 45] at (axis cs: 0.5, 0.90) {\scriptsize 1.50$\times$};
    \node[anchor=south west, rotate = 45] at (axis cs: 1.5, 1.47) {\scriptsize 2.07$\times$};
    \node[anchor=south west, rotate = 45] at (axis cs: 2.5, 3.52) {\scriptsize 4.12$\times$};
    \node[anchor=south west, rotate = 45] at (axis cs: 3.5, 3.75) {\scriptsize 4.35$\times$};
    \node[anchor=south west, rotate = 45] at (axis cs: 4.5, 3.75) {\scriptsize 4.35$\times$};

\end{axis}
    
\end{tikzpicture}

\vspace{-3mm}
\caption{From spare memory}
\label{fig:memory_sweep_a}
\end{subfigure}
\hfill
\begin{subfigure}[b]{0.48\linewidth}

\begin{tikzpicture}
\definecolor{TraditionalColor}{HTML}{ABDDA4}
\definecolor{MemcontainerColor}{HTML}{38A528}

\pgfplotstableread{
name
0.4
0.8
1.6
3.2
6.4
}\datatable

\pgfplotstableread{
key val
1 1.40
2 1.96
3 3.96
4 4.12
5 4.11
}\numbertable

\begin{axis}[
    ybar stacked,
                     legend style={ legend columns=4,
                                at={(-0.1,1.3)},anchor=south west,
                                column sep=0.1cm},  
    xtick=data,
    width=45mm,
    height=37mm,
    bar width=3mm,
    ymin=0,
    axis y line*=none,
    axis x line*=none,
    xticklabels from table={\datatable}{name},
    xtick={1, 2, 3, 4, 5},
    x tick label style={anchor=east,xshift=1.8ex, yshift = -1ex},
    ytick={1, 2, 3, 4, 5},
    yticklabels={1$\times$, 2$\times$, 3$\times$, 4$\times$, 5$\times$},
    ymax = 5,
    ymin = 1,
    xmax = 6,
    xmin = 0,
    tick label style={font=\footnotesize},
    legend style={font=\footnotesize,yshift=-6ex},
    label style={font=\footnotesize},
    ylabel={Speedup},
        y label style={at={(axis description cs:-0.2,.5)},anchor=south},
    xlabel={Memory ($\%$)},
    xlabel style={yshift = 1ex},
    area legend
    ]    
    

    \addplot [SCColor,fill=SCColor] table[x=key,y=val] {\numbertable};
    \node[anchor=south west, rotate = 45] at (axis cs: 0.5, 0.80) {\scriptsize 1.40$\times$};
    \node[anchor=south west, rotate = 45] at (axis cs: 1.5, 1.36) {\scriptsize 1.96$\times$};
    \node[anchor=south west, rotate = 45] at (axis cs: 2.5, 3.36) {\scriptsize 3.96$\times$};
    \node[anchor=south west, rotate = 45] at (axis cs: 3.5, 3.52) {\scriptsize 4.12$\times$};
    \node[anchor=south west, rotate = 45] at (axis cs: 4.5, 3.51) {\scriptsize 4.11$\times$};
    \end{axis}
\end{tikzpicture}

\vspace{-3mm}
\caption{From query memory}
\label{fig:memory_sweep_b}
\end{subfigure}
\caption{\system achieves more savings with larger \memory on the 100GB \tpcdsdate dataset. Significant savings are achieved even with a small \memory (0.4$\%$) relative to dataset size. \system's \memory can also be reallocated from query memory with minimal impact on speedup.
}
\label{fig:memory_sweep}
\vspace{-4mm}
\end{figure}

\subsection{\system Handles Large Workloads}
\label{sec:exp_size}

In this section, we evaluate the performance gain from \system across varying dataset 
sizes. We measure the speedup achieved by \system on the total execution time of the 5 workloads on regular and date-partitioned TPC-DS datasets of different sizes (10GB--1TB), with \memory 1.6$\%$ of the dataset size.

\Cref{fig:scale_sweep} report the end-to-end execution time reduction achieved by \system across different TPC-DS dataset scales. Given an appropriately sized \memory relative to dataset size, \system's optimization results in consistent execution time reductions across different dataset scales - 1.58$\times$--1.71$\times$ on the regular TPC-DS datasets, and 2.31$\times$--4.26$\times$ on the date-partitioned TPC-DS datasets.

\subsection{Efficient Usage of \memory}
\label{sec:exp_memory}

In this section, we study \system's usage of \memory by evaluating its performance gain with varying \memory sizes.
We measure speedup on both the total table read latency and total execution time of the 5 workloads on the 100GB \tpcdsdate dataset, performing a parameter sweep on \memory size from 0.4$\%$ to 6.4$\%$ of data size.
Additionally, we compare the achieved speedup between using spare memory in the system and reallocating query memory from the DBMS as \memory for \system. This is reported in \Cref{fig:memory_sweep}.

Storage of intermediate data in memory results in a notable speedup of 1.50$\times$ even with \memory 0.4$\%$ of data size.
Further increasing the \memory size results in a speedup of up to 4.26$\times$ with 6.4$\%$ of data size.
 
In the case where no spare memory is available in the system to act as \memory for \system, the \memory can also be reallocated from the query memory of the DBMS with minimal impact on achieved speedup. 
The reduction in speedup when using reassigned memory compared to spare memory is negligible - only up to $0.25\times$.

\subsection{\system Reduces Table Read Times}
\label{sec:exp_io}
In this section, we investigate \system's query-level benefits. We run the 5 workloads on 100GB datasets with varying \memory sizes and present CPU metrics reported by Presto.

We provide the breakdown of savings by \memory size in \cref{tbl:hardware}. The reduction of table read latency is a major source of \system's speedup, reaching 1.51$\times$ and 1.42$\times$ respectively with 6.4$\%$ of data size. \system's optimization has a negligible effect on compute latency, which is expected as it is not the target of \system's optimization.
 
 \begin{table}[t]
\vspace{2mm}
\caption{Effect of \system's optimization on various metrics on the 100GB datasets. \system reduces table read times by storing intermediate tables in the \memory.}
\addtolength{\tabcolsep}{-3pt} 
\begin{tabularx}{\columnwidth}{llllllll}
\toprule
Dataset & Latency(CPU,s) & No opt & 0.4$\%$ & 0.8$\%$ & 1.6$\%$ & 3.2$\%$ & 6.4$\%$ \\
\midrule
& Table read     & 4243 & 4308       & 3934 & 3574 & 3128 & 2884             \\
\tpcds & Compute     & 8533 & 8587       & 8319 & 8283 & 8249 &   8286           \\
 & Query     & 12776 & 12895       & 12253 & 11857 & 11377 &   11170           \\
\midrule
 & Table read     & 1710 & 1514       & 1314 & 1106 & 1106 & 1096             \\
\tpcdsdate & Compute     & 2843 & 2756       & 2709 & 2657 & 2636 &   2644           \\
 & Query     & 4553 & 4270       & 4023 & 3763 & 3742 &   3740           \\
\bottomrule
\end{tabularx}
\addtolength{\tabcolsep}{3pt}
\vspace{-4mm}
\label{tbl:hardware}
\end{table}

\subsection{Our Optimization
Finds Higher-Quality Solutions}
\label{sec:exp_algm}

In this section, we examine the effectiveness of our proposed solution (\solutionnodes + \solutionorder) to \optimizationproblem. We compare the time savings achieved by solving \optimizationproblem with \solutionnodes + \solutionorder to that of ablated methods---one of \solutionnodes or \solutionorder paired with an alternative method---under the same \memory size on the total execution time of the 5 workloads on the 100GB TPC-DS dataset.

\Cref{fig:algorithm} showcases this comparison on a variety of datasets. 
Our solutions to both subproblems---\solutionnodes for \subproblemnodes and \solutionorder for \subproblemorder---both outperform their respective alternative methods.
\solutionnodes paired with \solutionorder achieves up to 1.09$\times$ speedup compared to either \greedy, \random or \ratio paired with \solutionorder; similarly, \solutionorder paired with \solutionnodes achieves up to 1.21$\times$ speedup compared to either \simulatedannealing or \separator paired with \solutionnodes.

Notably, our individual solutions overcome pitfalls that affect the alternative methods:\begin{itemize}
    \item \greedy, \ratioshort, and \random flag nodes without considering the duration the node will be kept in memory.
    \item \simulatedannealing and \separator have poor compatibility with the \memory size constraint, which (i) limits the number of valid node swaps \simulatedannealing can perform and (ii) cannot be integrated into \separator commonly resulting in it unfeasible execution orders, ending alternating optimization early.
\end{itemize}

\subsection{\system Scales in Cluster-Based Environments}

\begin{figure}[t]

\begin{subfigure}[b]{0.48\linewidth}
\centering

\begin{tikzpicture}
\pgfplotstableread{ 
Method M1 M2 M3 M4 M5 M6 M7
\textbf{MKP+DFS'(ours)} 0 0 0 0 0 0 934.20
MKP+separator 0 0 0 0 0 1020.82 0
MKP+SA 0 0 0 0 1018.83 0 0
Ratio+DFS' 0 0 0 950.14 0 0 0
Greedy+DFS' 0 0 988.05 0 0 0 0
Random+DFS' 0 953.87 0 0 0 0 0
No+optimization 1528 0 0 0 0 0 0
}\datatable

\begin{axis}[
legend style={ legend columns=3,
                                at={(-0.2,1.1)},anchor=south west,
                                column sep=0mm},
    legend style={font=\scriptsize},
    width=45mm,
    height=28mm,
    bar width=1.5mm,
    xlabel={Time (s)},
    ylabel={Method},
    label style={font=\footnotesize},
    tick label style={font=\footnotesize},
    xlabel style={yshift = 1ex},
    axis y line*=none,
    axis x line*=none,
    xbar stacked,   
    xmin=800,
    xmax=1600,
    xtick={800, 1000, ..., 1600},
    ymin=-1,
    ytick=data,     
    yticklabels={,,},  
    /tikz/every even column/.append style={column sep=3mm},
]

    
\addplot [NoOptColor,fill=NoOptColor,x tick label style={xshift=-0.3cm}] table [x=M1, y expr=\coordindex] {\datatable};    
\addlegendentry[]{No Opt};

\addplot [RandomColor,fill=RandomColor,x tick label style={xshift=-0.3cm}]table [x=M2, y expr=\coordindex] {\datatable};
\addlegendentry[]{\random + \solutionorder};

\addplot [GreedyColor,fill=GreedyColor,x tick label style={xshift=-0.3cm}] table [x=M3,y expr=\coordindex] {\datatable};
\addlegendentry[]{\greedy + \solutionorder};

\addplot [HeuristicColor,fill=HeuristicColor,x tick label style={xshift=-0.3cm}] table [x=M4,y expr=\coordindex] {\datatable};
\addlegendentry[]{\ratioshort + \solutionorder};

\addplot [SAColor,fill=SAColor,x tick label style={xshift=-0.3cm}] table [x=M5,y expr=\coordindex] {\datatable};
\addlegendentry[]{\solutionnodes + \simulatedannealing};

\addplot [SeparatorColor,fill=SeparatorColor,x tick label style={xshift=-0.3cm}] table [x=M6,y expr=\coordindex] {\datatable};
\addlegendentry[]{\solutionnodes + \separator};

\addplot [black,fill=SCColor,x tick label style={xshift=-0.3cm}] table [x=M7,y expr=\coordindex] {\datatable};
\addlegendentry[]{\textbf{\solutionnodes + \solutionorder (Ours)}};

\end{axis}
    
\end{tikzpicture}
\vspace{-7mm}
\caption{\footnotesize \tpcds(1.6$\%$ \memory)}
\label{fig:algorithm_a}
\end{subfigure}
\hfill
\vspace{-1mm}
\begin{subfigure}[b]{0.48\linewidth}
\centering

\begin{tikzpicture}
\pgfplotstableread{ 
Method M1 M2 M3 M4 M5 M6 M7
\textbf{MKP+DFS'(ours)} 0 0 0 0 0 0 340.70
MKP+separator 0 0 0 0 0 420.48 0
MKP+SA 0 0 0 0 358.05 0 0
Ratio+DFS' 0 0 0 403.59 0 0 0
Random+DFS' 0 0 351.86 0 0 0 0
Greedy+DFS' 0 358.55 0 0 0 0 0
No+optimization 700 0 0 0 0 0 0
}\datatable

\begin{axis}[
legend style={ legend columns=7,
                                at={(-0,1.1)},anchor=south west,
                                column sep=0.1cm},
    legend style={font=\scriptsize},
    width=45mm,
    height=28mm,
    bar width=1.5mm,
    xlabel={Time (s)},
    ylabel={Method},
    label style={font=\footnotesize},
    tick label style={font=\footnotesize},
    xlabel style={yshift = 1ex},
    axis y line*=none,
    axis x line*=none,
    xbar stacked,   
    xmin=200,
    xmax=700,
    ymin=-1,
    ytick=data,     
    yticklabels={,,},  
]

    
\addplot [NoOptColor,fill=NoOptColor,x tick label style={xshift=-0.3cm}] table [x=M1, y expr=\coordindex] {\datatable};    

\addplot [RandomColor,fill=RandomColor,x tick label style={xshift=-0.3cm}]table [x=M2, y expr=\coordindex] {\datatable};

\addplot [GreedyColor,fill=GreedyColor,x tick label style={xshift=-0.3cm}] table [x=M3,y expr=\coordindex] {\datatable};

\addplot [HeuristicColor,fill=HeuristicColor,x tick label style={xshift=-0.3cm}] table [x=M4,y expr=\coordindex] {\datatable};

\addplot [SAColor,fill=SAColor,x tick label style={xshift=-0.3cm}] table [x=M5,y expr=\coordindex] {\datatable};

\addplot [SeparatorColor,fill=SeparatorColor,x tick label style={xshift=-0.3cm}] table [x=M6,y expr=\coordindex] {\datatable};

\addplot [black,fill=SCColor,x tick label style={xshift=-0.3cm}] table [x=M7,y expr=\coordindex] {\datatable};

\end{axis}
\end{tikzpicture}
\vspace{-2.3mm}
\caption{\footnotesize \tpcdsdate(.8$\%$ \memory)}
\label{fig:algorithm_b}
\end{subfigure}
\vspace{1mm}
\caption{
Comparison between different methods for solving \optimizationproblem on the 100GB datasets. 
Our solution (\solutionnodes + \solutionorder) outperforms ablated methods by saving up to an additional 3\%--11\% of workload execution time.
}
\vspace{-4mm}
\label{fig:algorithm}
\end{figure}
\label{sec:exp_distributed}
In this section, we showcase \system's ability to achieve speedup in a distributed setting.
We vary the worker node count (50GB query memory each) in our Presto cluster and investigate the relationship between node count and total end-to-end time speedup of the 5 workloads on the 100GB \tpcds dataset.

We present the results in \cref{tbl:distributed}. While the total end-to-end job execution time significantly decreases with each additional node in the distributed DB cluster, the additional savings achieved by \system's optimization remains largely consistent irrespective of the number of nodes.

\subsection{\system's Handling of Complex Workload Structures}
\label{sec:exp_complexity}

In this section, we study the effect of workload dependency structure on the optimization time and achievable speedup of \system. We use our workload generator to create DAGs/workloads of up to 100 nodes and (1) compare \system's optimization time with baseline methods and (2) perform sweeps to investigate the relationship between generation parameters and estimated savings from \system. We generate 1000 DAGs for each setting.

\mypara{Optimization Time} \Cref{fig:scalability} shows the average optimization time of the methods DAGs with up to 100 nodes.
The optimization time of our proposed solution, \solutionnodes + \solutionorder, scales linearly with the number of nodes in the DAG, with a negligible runtime of 0.02 seconds on DAGs with 100 nodes. 
While \greedy + \solutionorder, \random + \solutionorder, and \ratio + \solutionorder have faster optimization times, the advantage (0.001s/0.022s/0.008s vs 0.024s on 100-node graphs) is negligible compared to the additional end-to-end workload execution time reduction brought by \solutionnodes + \solutionorder (\Cref{sec:exp_algm}).
Both \solutionnodes + \simulatedannealing and \solutionnodes + \separator are significantly slower compared to the \solutionnodes + \solutionorder in terms of optimization time while also providing less benefit in terms of workload execution time reduction.
\begin{table}[t]
\caption{Effect of \system's optimization in DB clusters on the 100GB \tpcds dataset with 1.6$\%$ \memory.}
\addtolength{\tabcolsep}{-0.5pt} 
\begin{tabularx}{\columnwidth}{llllll}
\toprule
Metric & 1 node & 2 nodes & 3 nodes & 4 nodes & 5 nodes \\
\midrule
No opt runtime (s)     & 1528 & 868 & 656 & 546 & 487             \\
\system runtime (s)     & 934 & 521       & 383 & 333 & 304           \\
 Speedup    & 1.63$\times$ & 1.67$\times$       & 1.71$\times$ & 1.64$\times$ & 1.60$\times$    \\
\bottomrule
\end{tabularx}
\addtolength{\tabcolsep}{0.5pt}
\vspace{-2mm}
\label{tbl:distributed}
\end{table}
\subfile{plots/scalability_breakdown}

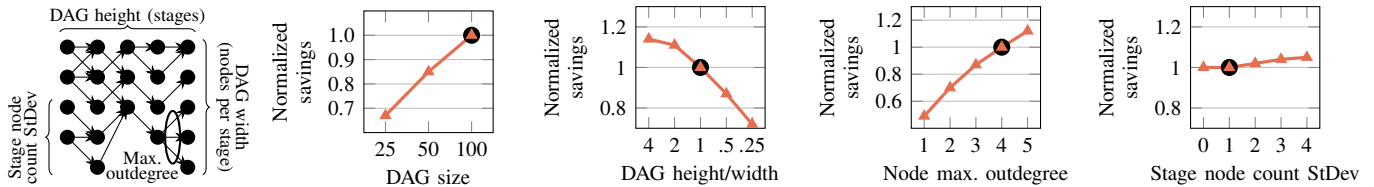
\begin{figure*}[t]

\pgfplotsset{accfig/.style={
    width=33mm,
    height=32mm,
    xmin=10,
    ymin=0,
    ymax=150,
    xlabel=Dependency graph size ({=} number of nodes),
    ylabel=RMS Error,
    xlabel near ticks,
    ylabel near ticks,
    ylabel style={align=center,font=\footnotesize\linespread{0.1},align=center},
    xtick={0, 10, ..., 100},
    ytick={0, 30, 60, 90, 120, 150},
    yticklabels={0\%, 30\%, 60\%, 90\%, 120\%, 150\%},
    ylabel shift=-2pt,
    xlabel shift=-2pt,
    legend style={
        at={(-0, 1.1)},anchor=south west,column sep=2pt,
        draw=black,fill=none,line width=.5pt,
        /tikz/every even column/.append style={column sep=3pt},
        font=\scriptsize
    },
    legend columns=8,
    every axis/.append style={font=\footnotesize},
}}

\pgfplotsset{timefig/.style={
    width=58mm,
    height=32mm,
    xmin=10,
    xmax=100,
    ymin=0,
    ymax=60,
    xlabel=Number of Queries,
    ylabel=Runtime (sec),
    xlabel near ticks,
    ylabel near ticks,
    ylabel style={align=center},
    xtick={0, 10, ..., 100},
    ytick={0, 10, ..., 60},
    ylabel shift=-2pt,
    xlabel shift=-2pt,
    legend style={
        at={(0, 1)},anchor=south west,column sep=2pt,
        draw=black,fill=none,line width=.5pt,
        /tikz/every even column/.append style={column sep=5pt}
    },
    legend columns=4,
    every axis/.append style={font=\footnotesize},
    minor grid style=lightgray,
}}
\begin{subfigure}[b]{0.16\linewidth}
\centering
\begin{tikzpicture}
\begin{scope}[every node/.style={circle, minimum width = 2mm, minimum height = 2mm,fill opacity = 1,fill=black, text opacity=1, inner sep = 0mm}]
    \node[] (X00) at (0, 1.6) {};
    \node[] (X01) at (0, 1.2) {};
    \node[] (X02) at (0, 0.8) {};
    \node[] (X03) at (0, 0.4) {};
    \node[] (X10) at (0.4, 1.6) {};
    \node[] (X11) at (0.4, 1.2) {};
    \node[] (X12) at (0.4, 0.8) {};
    \node[] (X13) at (0.4, 0.4) {};
    \node[] (X14) at (0.4, 0) {};
    \node[] (X20) at (0.8, 1.6) {};
    \node[] (X21) at (0.8, 1.2) {};
    \node[] (X22) at (0.8, 0.8) {};
    \node[] (X30) at (1.2, 1.6) {};
    \node[] (X31) at (1.2, 1.2) {};
    \node[] (X32) at (1.2, 0.8) {};
    \node[] (X33) at (1.2, 0.4) {};
    \node[] (X40) at (1.6, 1.6) {};
    \node[] (X41) at (1.6, 1.2) {};
    \node[] (X42) at (1.6, 0.8) {};
    \node[] (X43) at (1.6, 0.4) {};
    \node[] (X44) at (1.6, 0) {};
\end{scope}
\node[] (X55) at (2.7, 0) {};
\begin{scope}[>={Stealth[black]},
              every node/.style={fill=none,circle},
              every edge/.style={draw=black}]
    \draw[->] ($(X00)$) -- ($(X10)$);
    \draw[->] ($(X00)$) -- ($(X11)$);
    \draw[->] ($(X01)$) -- ($(X10)$);
    \draw[->] ($(X01)$) -- ($(X11)$);
    \draw[->] ($(X01)$) -- ($(X12)$);
    \draw[->] ($(X02)$) -- ($(X13)$);
    \draw[->] ($(X03)$) -- ($(X13)$);
    \draw[->] ($(X03)$) -- ($(X14)$);
    \draw[->] ($(X10)$) -- ($(X21)$);
    \draw[->] ($(X11)$) -- ($(X20)$);
    \draw[->] ($(X11)$) -- ($(X22)$);
    \draw[->] ($(X12)$) -- ($(X21)$);
    \draw[->] ($(X13)$) -- ($(X22)$);
    \draw[->] ($(X14)$) -- ($(X22)$);
    \draw[->] ($(X20)$) -- ($(X31)$);
    \draw[->] ($(X21)$) -- ($(X32)$);
    \draw[->] ($(X22)$) -- ($(X33)$);
    \draw[->] ($(X20)$) -- ($(X30)$);
    \draw[->] ($(X30)$) -- ($(X40)$);
    \draw[->] ($(X31)$) -- ($(X40)$);
    \draw[->] ($(X32)$) -- ($(X41)$);
    \draw[->] ($(X33)$) -- ($(X42)$);
    \draw[->] ($(X33)$) -- ($(X43)$);
    \draw[->] ($(X33)$) -- ($(X44)$);
\end{scope}

\begin{scope}[>={Stealth[black]},
              every node/.style={fill=none,circle},
              every edge/.style={draw=black}]
    \draw [decorate,
    decoration = {brace}] (-0.1, 1.8) --  (1.7, 1.8);
    \draw [decorate,
    decoration = {brace}] (1.8, 1.7) --  (1.8, -0.1);
    \draw [decorate,
    decoration = {brace}] (-0.2, -0.1) --  (-0.2, 0.9);
\end{scope}
\node[align=center] (X) at (0.8, 2.0) {\scriptsize DAG height (stages)};
\node[align=center, rotate=-90] (X) at (2.2, 0.8) {\scriptsize DAG width\\[-0.6em]\scriptsize (nodes per stage)};
\node[align=center, rotate=90] (X) at (-0.6, 0.4) {\scriptsize Stage node\\[-0.6em]\scriptsize count StDev};
\draw[thick] (1.4, 0.4) ellipse (1mm and 3.5mm);
\node[align=center] (X) at (1.0, 0) {\scriptsize Max.\\[-0.6em]\scriptsize outdegree};
\end{tikzpicture}
\end{subfigure}
\hspace{1mm}
\hfill
\begin{subfigure}[b]{0.20\linewidth}
\centering
\begin{tikzpicture}
\begin{axis}[accfig,
ylabel=Normalized\\savings,
ymin =0.6,
ymax=1.1,
xmin = 0.5,
xmax = 3.5,
xlabel=DAG size,
xtick={1, 2, 3},
xticklabels={25, 50, 100},
ytick={0.7, 0.8, 0.9, 1.0},
yticklabels={0.7, 0.8, 0.9, 1.0},
ymajorgrids,
    ]

\addplot[mark=triangle*,mark size=1.5,mark options={fill=SCColor,draw=SCColor},
very thick,draw=SCColor,
]
table[x=x,y=y] {
x y
1 0.67
2 0.85
3 1.00
};

\node (mark) [draw, draw=black, fill=black, circle, minimum size = 5pt, inner sep=2pt, thick] 
      at (axis cs: 3, 1.00) {};

\end{axis}
\end{tikzpicture}
\end{subfigure}
\hfill
\begin{subfigure}[b]{0.20\linewidth}
\centering
\begin{tikzpicture}
\begin{axis}[accfig,
ylabel=Normalized\\savings,
ymin =0.7,
ymax=1.3,
xmin = 0.5,
xmax = 5.5,
xlabel=DAG height/width,
xtick={1, 2, 3, 4, 5},
xticklabels={4, 2, 1, .5, .25},
ytick={0.8,1,1.2},
yticklabels={0.8,1,1.2},
ymajorgrids,
    ]

\addplot[mark=triangle*,mark size=1.5,mark options={fill=SCColor,draw=SCColor},
very thick,draw=SCColor,
]
table[x=x,y=y] {
x y
1 1.14
2 1.11
3 1.00
4 0.87
5 0.72
};

\node (mark) [draw, draw=black, fill=black, circle, minimum size = 5pt, inner sep=2pt, thick] 
      at (axis cs: 3, 1.00) {};

\end{axis}
\end{tikzpicture}
\end{subfigure}
\hfill
\begin{subfigure}[b]{0.20\linewidth}
\centering
\begin{tikzpicture}
\begin{axis}[accfig,
ylabel=Normalized\\savings,
ymin =0.4,
ymax=1.3,
xmin = 0.5,
xmax = 5.5,
xlabel=Node max. outdegree,
xtick={1, 2, 3, 4, 5},
xticklabels={1, 2, 3, 4, 5},
ytick={0.6, 0.8,1,1.2},
yticklabels={0.6, 0.8,1,1.2},
ymajorgrids,
    ]

\addplot[mark=triangle*,mark size=1.5,mark options={fill=SCColor,draw=SCColor},
very thick,draw=SCColor,
]
table[x=x,y=y] {
x y
1 0.49
2 0.70
3 0.87
4 1.00
5 1.12
};

\node (mark) [draw, draw=black, fill=black, circle, minimum size = 5pt, inner sep=2pt, thick] 
      at (axis cs: 4, 1.00) {};

\end{axis}
\end{tikzpicture}
\end{subfigure}
\hfill
\begin{subfigure}[b]{0.20\linewidth}
\centering
\begin{tikzpicture}
\begin{axis}[accfig,
ylabel=Normalized\\savings,
ymin =0.7,
ymax=1.3,
xmin = 0.5,
xmax = 5.5,
xlabel=Stage node count StDev,
xtick={1, 2, 3, 4, 5},
xticklabels={0, 1, 2, 3, 4},
ytick={0.8,1,1.2},
yticklabels={0.8,1,1.2},
ymajorgrids,
    ]

\addplot[mark=triangle*,mark size=1.5,mark options={fill=SCColor,draw=SCColor},
very thick,draw=SCColor,
]
table[x=x,y=y] {
x y
1  1.00
2  1.00
3  1.02
4  1.04
5  1.05
};
\node (mark) [draw, draw=black, fill=black, circle, minimum size = 5pt, inner sep=2pt, thick] 
      at (axis cs: 2, 1.00) {};

\end{axis}
\end{tikzpicture}
\end{subfigure}

\vspace{-2mm}
\caption{
Analysis of DAG complexity vs. predicted savings from \system on synthetic DAGs. Savings normalized w.r.t. DAGs with 100 nodes, height/width ratio 1, max. out-degree 4, and stage node count StDev 1 (Parameters in \cref{fig:scalability}, marked in black).
}
\vspace{-4mm}
\label{fig:complexity}
\end{figure*}
\mypara{Parameter Sweep}
\Cref{fig:complexity} reports the results of \system's optimization vs. various generation parameters. 
Time savings are normalized w.r.t. parameters in \cref{fig:scalability} marked in black.
\begin{itemize}
    \item \textbf{DAG size:} The normalized savings achieved by \system is highly correlated with DAG size. The relationship is not proportional; however, this is expected as deeply nested MVs tend to be smaller from repeated filters/projections.
    \item \textbf{DAG height/width:} The workload generator creates DAGs following the structure of Spark workloads:
    The height and width correspond to the number of stages in the workload and the number of nodes per stage, respectively.
    \system achieves more savings on 'thinner' DAGs with higher height/width ratio due to fewer inter-stage dependencies, allowing nodes kept in \memory to be freed sooner with an efficient execution order.
    \item \textbf{Node max. outdegree:} This parameter controls the number of outgoing edges generated for each node in the DAG, uniformly sampled from [0, max. outdegree].
    A higher value results in more savings, a result of individual nodes having higher speedup scores as its flagging reduces the file read latency of more downstream nodes.
    \item \textbf{Stage node count standard deviation (StDev):} This parameter introduces variance into the number of nodes per stage in the DAG. While the DAG structure becomes increasingly irregular with higher variance, the irregularity's effect on speedup is negligible.
\end{itemize}

\section{Related work}

 
\mypara{Efficient MV Refresh}
Incremental view maintenance (IVM) aims to update MVs to reflect newly ingested data, taking advantage of already computed results to perform the update in a manner more efficient than computing from scratch (full refresh)
~\cite{ahmad2012dbtoaster,mcsherry2013differential,armbrust2013generalized,zeng2016iolap, palpanas2002incremental, griffin1995incremental, agiwal2021napa, braun2015analytics}. 
There is an abundance of work in IVM, including incremental updates on duplicate values~\cite{griffin1995incremental}, non-distributive aggregate functions~\cite{palpanas2002incremental}, higher-order views~\cite{ahmad2012dbtoaster}, and sliding windows~\cite{braun2015analytics}. 
More recent works also investigate the scalability aspect of IVM, proposing scale-independent updates~\cite{armbrust2013generalized} and sampled views~\cite{zeng2016iolap}. Since \system is applicable to arbitrary SQL statements, \system is orthogonal to and is fully compatible with existing IVM techniques.

\mypara{MV Refresh Scheduling}
There exist works on scheduling the refresh of a MV set focusing on resolving cyclic dependencies~\cite{folkert2005optimizing}, minimizing weighted average staleness~\cite{golab2009scheduling}, and prioritizing MVs with the highest speedups on predicted future queries~\cite{ahmed2020automated}.
\system's scheduling to speed up the end-to-end refresh of the MV set is not addressed in existing works.

\mypara{DAG Workflow Scheduling}
The execution of workloads consisting of individual jobs with acyclic dependencies is a well-studied topic~\cite{apacheoozie,sparkdag,marchal2018parallel,bathie2020revisiting,baruah2022ilp}; many of these techniques can be applied to MV refresh runs studied in this paper.
Existing workflow scheduling systems such as Apache Oozie~\cite{apacheoozie}, Apache Airflow~\cite{airflow}, and Spark DAG scheduler~\cite{sparkdag} automate the execution of user-defined workflows following a topological order.
There are recent works aimed at finding more optimal execution orders in terms of peak memory usage~\cite{marchal2018parallel, bathie2020revisiting} and execution time on parallel platforms~\cite{baruah2022ilp}.
While \system is designed for use with MV refresh runs/workloads, our technique on joint scheduling and optimization can be reasonably applied to general workloads as a possible future direction.

\mypara{Intermediate Data Caching}
Some existing data visualization systems cache user-defined variables to support the typical incremental construction of data visualizations~\cite{zgraggen2016progressive, eichmann2020idebench} during data analysis sessions~\cite{jupyter, rstudio, colab}. 
Recent work proposes a management scheme for these cached variables under a memory constraint that greedily keeps variables with the highest estimated time savings based on predicted future user behavior via neural networks~\cite{xin2021enhancing}.
While useful for data visualization, a greedy approach to memory management fails to achieve satisfactory results compared to \system.

\mypara{Intermediate Result Reuse}

There exist works on storing intermediate results from computations to speedup future computations~\cite{yang2018intermediate, dursun2017revisiting, nagel2013recycling, michiardi2019memory, galakatos2017revisiting}.
Studied topics include the identification of reuse opportunities by finding overlaps in computation graphs of successive jobs~\cite{yang2018intermediate, michiardi2019memory},
selective storage under a space constraint with heuristics such as reuse probability~\cite{dursun2017revisiting}, expected savings~\cite{yang2018intermediate}, and recompute-storage cost difference~\cite{nagel2013recycling},
and rewriting incoming jobs to take advantage of stored intermediates~\cite{galakatos2017revisiting}.
These works share similarity with \system in their selection of items to store under a memory constraint, however, \system's problem setting requires it to uniquely consider the joint (re)ordering of job executions along with the selection of items.


\mypara{Incremental Query Processing} Incremental processing (IQP) is useful for cases where not all data required for a query is immediately available. Similar to online aggregation~\cite{hellerstein1997online}, initial results of a query are computed on a subset of required data and progressively refined as the rest of the required data arrives in a predictable pattern~\cite{tang2019intermittent,wangtempura}. Tang et al. propose a dynamic programming formulation to pick intermediate states to store in memory given a limited memory budget~\cite{tang2019intermittent}. Tempura rewrites the query plan for more efficient execution based on predicted data arrival patterns~\cite{wangtempura}. While similarities exist between the problem setting of IQP and \system, such as management of bounded memory, \system notably includes additional joint optimization for the order of MV updates.


\section{Conclusion and Future Work} 

Sequentially refreshing materialized views (MVs) 
    according to a topological order bears a noticeable overhead 
    as a result of the significant portion of intermediate I/O
        spent on reading and writing intermediate data.
In this paper, we propose a new system, \system, which aims to speed up MV refresh workloads 
    by selectively persisting intermediate data in bounded memory 
        during workload execution 
    to reduce the blocking I/O costs,
        thereby significantly reducing the
    wait times for reading and writing intermediate data from and to external storage.
We propose an effective and scalable algorithm
    to jointly optimize (1) a subset of intermediate data to persist and 
        (2) a topological order,
    in order to minimize the end-to-end MV refresh time under bounded memory.
\system is different from and complements existing MV refresh techniques as our approach is oblivious to individual MV refresh operations.
We demonstrate that \system can reduce end-to-end MV refresh workload execution times by 1.04$\times$--5.08$\times$ with 1.6GB memory on a 100GB dataset.
Moreover, our optimization algorithm can easily handle complex workloads with 100 MVs.

In the future, we plan to generalize techniques presented in this paper to non-MV refresh recurring workloads containing individual jobs with acyclic dependencies.


\section*{Acknowledgement}
This work is supported in part by Microsoft Azure.

\bibliographystyle{IEEEtran}
\bibliography{sample}

\end{document}